%% file: main_arxiv.tex
\documentclass[aps,superscriptaddress,nofootinbib,twocolumn,truncate=10]{revtex4-2}
\usepackage{bm}
\usepackage{lmodern}

\usepackage[utf8]{inputenc}
\usepackage{amsmath}
\usepackage{amssymb}
\usepackage{upgreek}
\usepackage{graphicx}
\usepackage{epstopdf}
\usepackage[dvipsnames]{xcolor}
\usepackage{enumerate}
\usepackage{gensymb}
\usepackage{ulem}
\usepackage[T1]{fontenc}
\newcommand{\stkout}[1]{\ifmmode\text{\sout{\ensuremath{#1}}}\else\sout{#1}\fi}
\usepackage[english]{babel}
\usepackage{braket}
\usepackage{float}
\usepackage{stmaryrd}
\usepackage{mathtools}
\usepackage[misc]{ifsym}
\usepackage{siunitx}
\usepackage{stackengine}
\usepackage{ctable}
\usepackage{makecell}

\usepackage{lineno}
\modulolinenumbers[5]

\usepackage{scalerel}

\def\vecsign#1{\rule[1.388\LMex]{\dimexpr#1-2.5pt}{.36\LMpt}%
  \kern-6.0\LMpt\mathchar"017E}

\stackMath
\usepackage{graphicx}

\newcommand{\AI}[1]{\textcolor{black}{#1}}

\newcommand\blankfootnote[1]{%
  \let\thefootnote\relax\footnotetext{#1}%
  \let\thefootnote\svthefootnote%
}

\newenvironment{myfont}{\fontfamily{ptm}\selectfont}{\par}
\DeclareTextFontCommand{\textmyfont}{\myfont}

\usepackage{hyperref}
\usepackage{lipsum}

\begin{document}

\title{Sparse qubit operation in a 6$\times$6 quantum dot array} 

\author{Alexander S. Ivlev}
\thanks{}
\affiliation{QuTech and Kavli Institute of Nanoscience, Delft University of Technology, PO Box 5046, 2600 GA Delft, The Netherlands}
\author{Harold L. Co}
\affiliation{QuTech and Kavli Institute of Nanoscience, Delft University of Technology, PO Box 5046, 2600 GA Delft, The Netherlands}
\author{Damien R. Crielaard}
\affiliation{QuTech and Kavli Institute of Nanoscience, Delft University of Technology, PO Box 5046, 2600 GA Delft, The Netherlands}
\author{Alice Petrillo}
\affiliation{QuTech and Kavli Institute of Nanoscience, Delft University of Technology, PO Box 5046, 2600 GA Delft, The Netherlands}
\author{Setareh Kazemzadeh}
\affiliation{QuTech and Kavli Institute of Nanoscience, Delft University of Technology, PO Box 5046, 2600 GA Delft, The Netherlands}
\author{Giordano Scappucci}
\affiliation{QuTech and Kavli Institute of Nanoscience, Delft University of Technology, PO Box 5046, 2600 GA Delft, The Netherlands}
\author{Menno Veldhorst}
\affiliation{QuTech and Kavli Institute of Nanoscience, Delft University of Technology, PO Box 5046, 2600 GA Delft, The Netherlands}

\date{\today}

\begin{abstract}
Fault-tolerant quantum computation requires increasing the qubit count while maintaining high performance and connectivity~\cite{dennisTopologicalQuantumMemory2002}. Semiconductor qubits have demonstrated high-fidelity operation~\cite{mills_2022_ScienceAdv_TwoqubitSiliconQuantum,Noiri_2022_Nature_Threshold,Xue_2022_Nature_SurfaceCodeThreshold,Wang_2024_Science_OperatingSemiconductorQuantum}, but growing the system size at sufficient connectivity remains challenging~\cite{franke_2019_Microprocessors&Systems_RentsRuleExtensibility}, such that scaling mainly occurred in one spatial direction to preserve access for control lines in the other~\cite{george12SpinQubitArraysFabricated2025,membersofthehrlquantumteamDigitallyControlledSilicon2026,dijkemaSimultaneousOperation18qubit2026}. 
Here, we realise an approach that relaxes the requirements on fabrication \cite{ivlevOperatingSemiconductorQubits2025a} and demonstrate a 6$\times$6 quantum dot array, defined by a shared barrier gate and individual plunger gates. The quantum dots are tuned through coherent spin shuttling over distances up to \SI{5.2}{\micro m} into a sparse configuration of 10 simultaneously operated qubits, leaving space for shuttling connections. We \AI{reduce crosstalk by $42\times$ in this separated regime}, enabling high-fidelity simultaneous single-qubit control, while control-Z operations can be implemented through shuttling and plunger-only control. These results promise high-connectivity quantum circuits, enabled by scaling to two-dimensional geometries and exploiting shuttling. 


\end{abstract}

\maketitle
Semiconductor spin qubits in gate-defined quantum dots have advanced to realisations of high-fidelity control in few-qubit systems~\cite{mills_2022_ScienceAdv_TwoqubitSiliconQuantum,Noiri_2022_Nature_Threshold,Xue_2022_Nature_SurfaceCodeThreshold,Wang_2024_Science_OperatingSemiconductorQuantum} and primitive error correcting codes ~\cite{vanriggelenPhaseFlipCode2022, takedaQuantumErrorCorrection2022, undsethWeightfourParityChecks2026,membersofthehrlquantumteamDigitallyControlledSilicon2026}. A key challenge toward the construction of larger systems is set by the interconnections, as the small size of quantum dots may result in a wiring bottleneck \cite{franke_2019_Microprocessors&Systems_RentsRuleExtensibility,vandersypenInterfacingSpinQubits2017}. As a result, the quantum dot count in experimental systems has mainly increased by expanding the system in a single direction, to make the fabrication feasible~\cite{membersofthehrlquantumteamDigitallyControlledSilicon2026,dijkemaSimultaneousOperation18qubit2026,george12SpinQubitArraysFabricated2025}.\\


Following this experimental restriction, a variety of architectural proposals have emerged relying on quasi one-dimensional quantum dot arrays, often connected through spin-shuttling links~\cite{kunne_2024_NatureComm_SpinBusArchitectureScaling,siegelEarlyFaultTolerance2024a,siegelSnakesPlaneMobile2025,liTrilinearQuantumDot2026}. These approaches create the effective two-dimensional topology necessary for fault-tolerant quantum computing, but \AI{currently} come with an overhead in time, space and circuit depth, \AI{resulting in stringent fidelity requirements}. These locally 1D spin qubit architectures stand in contrast with other qubit modalities where 2D connectivity is utilised directly, \AI{allowing for} efficient parallel syndrome extraction and below-threshold error correction cycles~\cite{acharyaQuantumErrorCorrection2025,bluvstein_2024_Nature_LogicalQuantumProcessor}. \AI{Applying such extended 2D geometries to spin qubits would both empower existing architectural proposals and unlock novel approaches, and a key challenge is then to realize the control for qubit operation~\cite{Borsoi_2024_NatureNano_SharedControl16,liCrossbarNetworkSilicon2018}.}\\

Here, we \AI{address} scaling in two dimensions while maintaining individual control, by combining a merged barrier approach~\cite{ivlevOperatingSemiconductorQubits2025a} with spin shuttling. By positioning \AI{hole} qubits sparsely in a 6 by 6 quantum dot array, we alleviate spurious exchange-interaction and crosstalk, enabling simultaneous single qubit control with above 99\% Clifford fidelities. Finally, by shuttling qubits into proximity and using plunger gates to control the exchange energy~\cite{ivlevOperatingSemiconductorQubits2025a}, we turn on the two-qubit interaction and realise controlled-Z gates between several qubits, demonstrating the foundational elements for quantum computation in sparsely occupied 2D quantum dot arrays.\\

\begin{figure*}
    \centering
    \includegraphics[width=\textwidth]{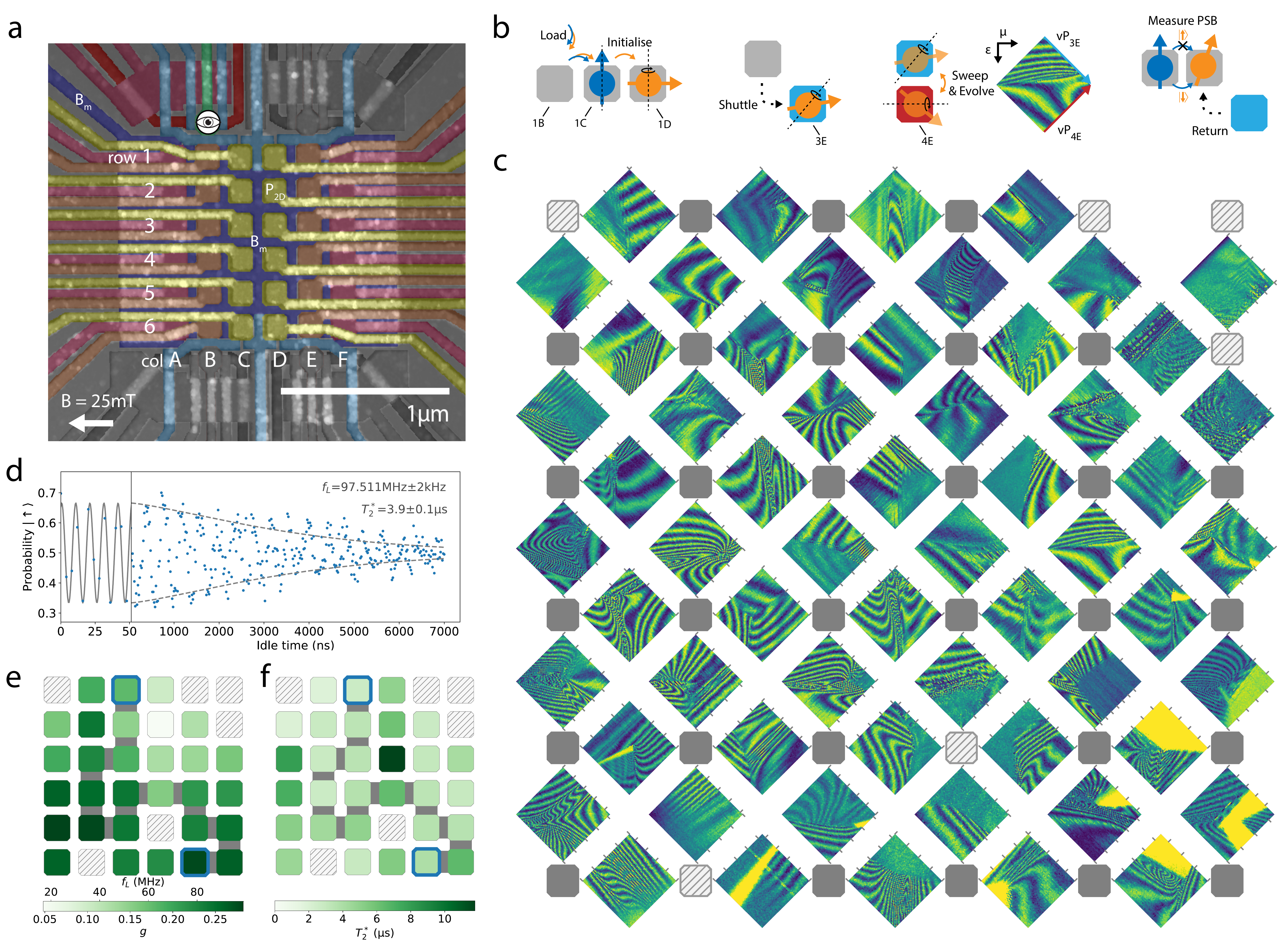}
    \caption{\textbf{A 6$\times$6 quantum dot array.} \textbf{a,} A false-colour Atomic Force Microscope image of the studied device. Plunger gates and quantum dots are labelled, e.g. $\text{P}_{2\text{D}}$. The merged barrier gate $\text{B}_{\text{m}}$ is indicated between the plunger gates and at its fanout (see Supplementary Material~\ref{supp:fabrication} for further details on the device). \AI{The direction of the nominally in-plane magnetic field $B=\SI{25}{mT}$ is indicated.} \textbf{b,} Schematic of a Spin Evolution Diagram (SED) measurement. A spin is loaded, prepared in a superposition state, and shuttled through the array to probe spatial variations in the g-factor. The accumulated phase is measured using PSB, yielding the example SED as a function of the voltages vP\textsubscript{3E}-vP\textsubscript{4E}. \textbf{c,} An overview of all SEDs obtained with a phase-accumulation time of \SI{500}{ns} throughout the array, oriented such that the corner near a quantum dot corresponds to the region in which the spin is underneath it. Hatched locations are not coherently accessible. The ticks indicate \SI{50}{mV} increments on the virtual plunger gate voltage. \textbf{d,} A time-resolved measurement of a spin from QD\textsubscript{1C}, idling in a distant QD\textsubscript{6E}. The path taken in this example is highlighted in \textbf{e,f} grey and consists of 26 dot-to-dot transitions, and a total of \SI{5.2}{\micro m} traversed. The g-factor and coherence time $T_2^*$ are extracted using a fitted decaying oscillation. For clarity, a subset of the data-points after \SI{50}{ns} is displayed and only the envelope of the fit is plotted. This is repeated throughout the array (\textbf{e}-\textbf{f}).}
    \label{fig:figure1}
\end{figure*}

We employ a merged-barrier design~\cite{ivlevOperatingSemiconductorQubits2025a} to realise a 6$\times$6 quantum dot array (Fig.~\ref{fig:figure1}a) \AI{in a Ge/SiGe heterostructure~\cite{lodari_MatQuantTech_LowPercolationDensity2021}.} The plunger gate voltages are used to control the tunnel coupling between neighbouring quantum dots, which aids in shuttling, two-qubit interaction, as well as spin initialisation and readout (see Supplementary Material~\ref{supp:fabrication} and \ref{supp:EarlierDevices} for device details and earlier designs). While the voltages on the merged barrier grid itself could be changed dynamically, this has a global effect throughout the array and is left unexplored in this work. Throughout all presented experiments, the voltage on the merged barrier grid and all other barrier gates is fixed. We note that the DC voltages on all plunger gates have also been kept constant since the initial tune-up, and any voltage drift has been compensated with AWG pulses~(see Supplementary Material~\ref{supp:ExperimentalSequence} for a discussion on voltage drift).\\

To bring the quantum dots in the array to the desired hole occupation, the voltages of the individual plunger gates need to be tuned. For the quantum dots far from the charge sensor, we can not rely on direct charge sensing, but instead develop an indirect charge sensing technique using a spin-qubit as a probe. First, two hole spin qubits are initialised close to the sensor, in typical fashion~\cite{hendrickx_2021_Nature_FourqubitGermaniumQuantum,Wang_2024_Science_OperatingSemiconductorQuantum,johnRobustLocalisedControl2025}~(see Fig.~\ref{fig:figure1}b and Supplementary Material~\ref{supp:ExperimentalSequence} for more details). Once one of the spins is brought into a superposition state using Electric Dipole Spin Resonance driving~(EDSR)~\cite{pioro-ladriereElectricallyDrivenSingleelectron2008}, it undergoes g-tensor-dependent Larmor precession. Crucially, this g-tensor is highly sensitive to electric fields~\cite{katoGigahertzElectronSpin2003,crippaElectricalSpinDriving2018,hendrickxFastTwoqubitLogic2020} and strain~\cite{abadillo-urielHoleSpinDrivingStrainInduced2023,martinezDisorderindependentHoleSpin2026}. Consequently, the Larmor precession will have a site-dependent voltage-susceptibility, resulting in a unique relation between the qubit phase and the plunger gate voltages as the spin is moved between different quantum dots and idles there~(Figure~\ref{fig:figure1}b,c). This phase is projected onto the Z-axis in the initial quantum dot pair and read out through latching-enhanced PSB~\cite{onoCurrentRectificationPauli2002,harvey-collardHighFidelitySingleShotReadout2018}. From the phase-information of the spin-evolution at different plunger gate voltages, the various charge-occupation regions can be determined, in what we call Spin Evolution Diagrams (SED).\\

In Figure~\ref{fig:figure1}b we see an example of one such SED, where changing the virtual~\cite{hensgens_2017_Nature_QuantumSimulationFermi} plunger gate voltages $\text{vP}_{3\text{E}}$ and $\text{vP}_{4\text{E}}$ produces two qualitatively distinct regions. We assign these regions to the different physical locations of the probing spin, separated by an extended interdot transition. In the sparse regime the hole is isolated, which results in a large voltage window over which the spin can coherently transition to the next site. This makes the charge shuttling resilient to moderate voltage drift, suitable for autonomous exploration~\cite{zwolakColloquiumAdvancesAutomation2023a}, and amenable to operate without precise virtual gate calibration in the future. Detailed simulations of more SED features are included in Supplementary Material~\ref{supp:SED}, where also decoherence and path-dependent effects are discussed. 
\\

Once the desired interdot transition is identified, the spin is shuttled in a bucket-brigade manner~\cite{fujitaCoherentShuttleElectronspin2017,millsShuttlingSingleCharge2019,yonedaCoherentSpinQubit2021,vanriggelendoelman_2024_NatureCom_CoherentSpinQubit} across that transition and the process is repeated for the next site. We find this indirect charge sensing and tuning through SEDs to be more reliable and \AI{more} interpretable compared to cascade readout and direct charge sensing (see Supplementary Material~\ref{supp:EarlierDevices} and~\ref{supp:incoherent_sites} for a discussion on these readout techniques). \\
\begin{figure*}
    \centering
    \includegraphics[width=\linewidth]{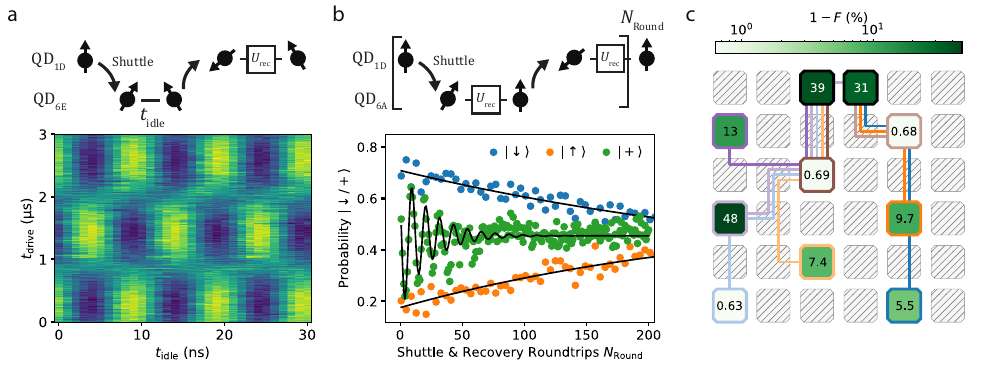}
    \caption{
    \textbf{Shuttling-based initialisation at sparse occupation.}
    \textbf{a,} Calibration of $t_{\text{drive}}$ for the return recovery gate $U_{\text{rec,6E-1D}}$ at a given microwave phase. $t_{\text{drive}}\approx\SI{0.9}{\micro s}$ is chosen as it minimises the spin dynamics as a function of the idle time $t_{\text{idle}}$. For further details on the exact calibration, see Supplementary Materials~\ref{supp:recovery_gates}. \textbf{c,}  For shuttling path QD\textsubscript{1C}-QD\textsubscript{6A} a basis-state round-trip fidelity of $F_{\text{Round}\ket{\downarrow}}=99.4\pm0.1\%$ is observed. For the superposition state an oscillatory signal as a function of the round trips $N_{\text{Round}}$ is observed, as no single-qubit phase corrections are calibrated. Its round-trip fidelity is fitted to $F_{\text{Round}\ket{+}}=95.7\pm0.5\%$. In addition to two recovery gates, this particular \SI{2.8}{\micro m} path consists of 14 bucket-brigade transitions giving lower-bound average fidelities per transition of $F_{\text{BB}\ket{\downarrow}}=99.95\%$ and $F_{\text{BB}\ket{+}}=99.7\%$ for a basis state and superposition state respectively.
    The basis-state transfer error is also extracted for the other 7 destination sites in \textbf{d}, following the coloured shuttling paths~(see Supplementary Material~\ref{supp:ExperimentalSequence} for details on shuttling order). For QD\textsubscript{1C} and QD\textsubscript{1D}, the error gives the average SPAM visibility of the spins originating from those sites.
    }
    \label{fig:figure3}
\end{figure*}

The overview of all Spin Evolution Diagrams used to tune the device is given in Figure~\ref{fig:figure1}c. We reach 30 out of 36 quantum dots coherently (see Supplementary Material~\ref{supp:incoherent_sites} for charge sensing of the other quantum dots and a discussion on coherence). By studying the free-induction decay, the g-factors and coherence time across the array are extracted. See for example Figure~\ref{fig:figure1}d, where the probe-spin from QD\textsubscript{1C} is moved back and forth across 14 distinct quantum dots, traversing a physical distance of \SI{2.6}{\micro m}, \AI{further than} even one-dimensional \AI{spin shuttling} experiments~\cite{ademiDistributingEntanglementDistant2026,zwerver_2023_PRX_ShuttlingElectronSpin,desmetHighfidelitySinglespinShuttling2025,undsethWeightfourParityChecks2026,volmerMappingValleySplitting2024,vanriggelendoelman_2024_NatureCom_CoherentSpinQubit}. The path is visualised in Figure~\ref{fig:figure1}e and f, where also g-factor and coherence properties for the other quantum dots throughout the array are displayed. The ability to measure these qubit properties across a 2D area at quantum dot resolution, uniquely adds to the existing metrology toolbox~\cite{volmerMappingValleySplitting2024,desmetHighfidelitySinglespinShuttling2025,ademiDistributingEntanglementDistant2026,tosatoCrossbarChipBenchmarking2026}, showing that the merged barrier provides a compelling platform for device characterization.\\ 

To prepare for qubit operation, basis states are initialised in the array by shuttling them along an explored path. After a spin is shuttled out to the target location, as well as on its return, a recovery gate~\cite{ademiDistributingEntanglementDistant2026} may be desired to compensate for shuttling-induced rotations resulting from site-dependent quantisation axes~\cite{vanriggelendoelman_2024_NatureCom_CoherentSpinQubit,Wang_2024_Science_OperatingSemiconductorQuantum,ademiDistributingEntanglementDistant2026,unseldBasebandControlSingleelectron2024}. In order to disentangle the required outbound and return recovery gates, we study the spin-evolution at the target quantum dot. If the outbound recovery gate is calibrated properly, the spin is initialised in an eigenstate of the target quantum dot, and no evolution should be observed. Similarly, if the return recovery gate is calibrated, the state is read out along the Z-basis during PSB, and no phase-accumulation \AI{should be visibile if a superposition is prepared in the target dot}. Since it enables phase-independent readout, the return recovery pulse is calibrated first. Figure~\ref{fig:figure3}a shows an example of this calibration for QD\textsubscript{6A}, giving a gate-time of about $t_{\text{drive}}\approx\SI{0.9}{\micro s}$. Details on the frequency and phase calibration, as well as the exact procedure can be found in Supplementary Material~\ref{supp:recovery_gates}.\\
\begin{figure*}
    \centering
    \includegraphics[width=0.97\linewidth]{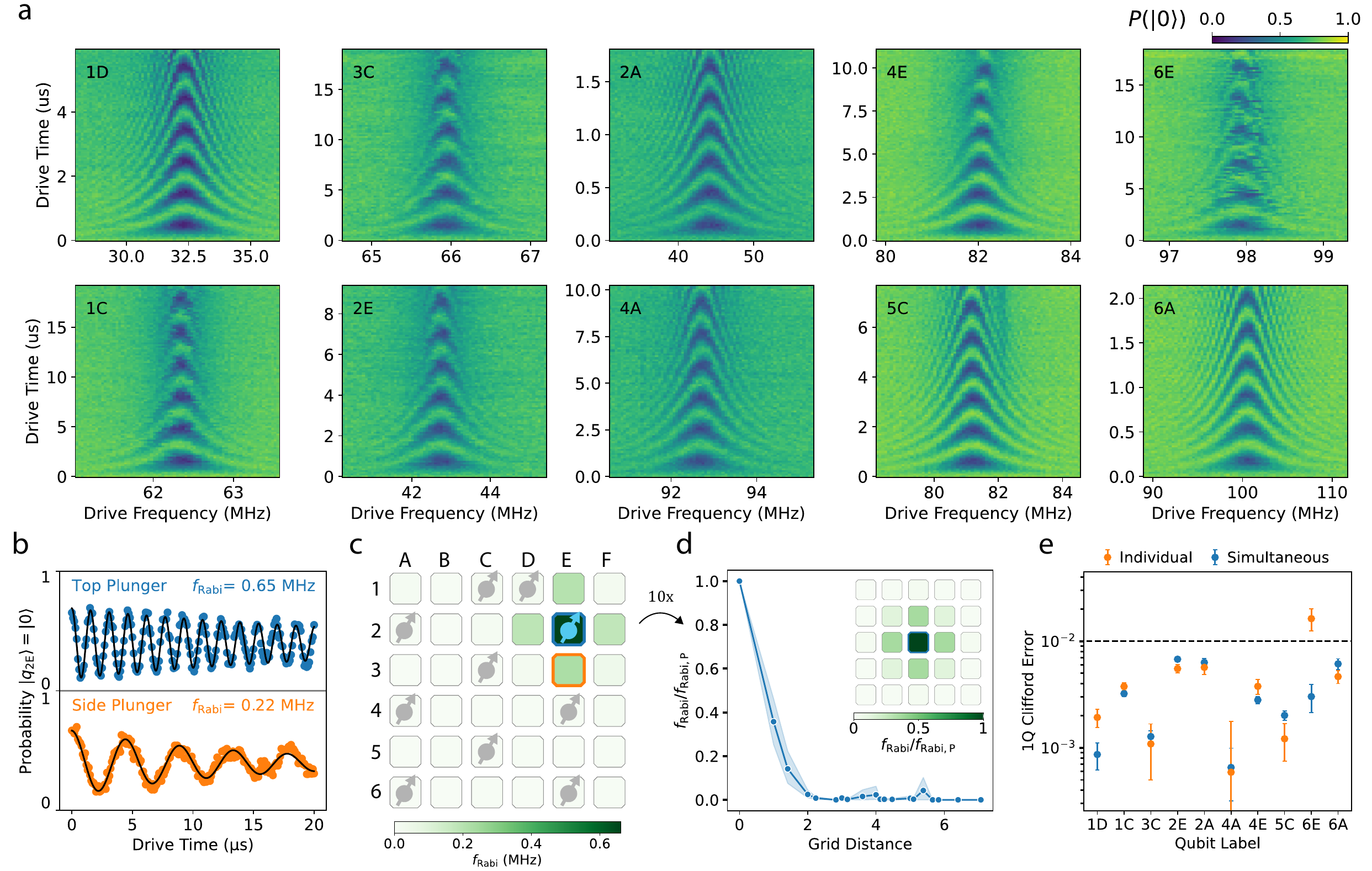}
    \caption{\textbf{Sparse single qubit operation}. \textbf{a,} Chevron plots of all ten qubits. The microwave driving is performed simultaneously per row. \textbf{b,} Qubit $q$\textsubscript{2E} is more efficiently driven with the top-plunger P\textsubscript{2E} (blue) than a side-plunger P\textsubscript{3E} (orange). These plungers are also highlighted in \textbf{c}, where an overview of the Rabi frequencies of $q$\textsubscript{2E} (cyan) is given when driving with any plunger. Remaining qubits are greyed out. \textbf{d,} Measuring the relative driving efficiencies of each plunger for the ten qubits, an average driving efficiency is calculated as a function of plunger-qubit distance. The shaded areas give the 95\% confidence interval. This average is normalised to the Rabi frequency of each top-plunger $f_\text{Rabi,P}$, and further visualised in the inset. \textbf{e,} Single-qubit Clifford error, as calculated through Randomised Benchmarking (RB). For both individual and simultaneous driving, single qubit errors typically fall within the surface code threshold of about 1\%~\cite{fowlerHighthresholdUniversalQuantum2009}. The details on the RB experiment can be found in the Methods.
    }
    \label{fig:figure4}
\end{figure*}

Once the recovery gates are calibrated in both directions we repeat the recovered shuttle sequence and measure the state decay to determine the round-trip fidelity. Figure~\ref{fig:figure3}b gives the decay curve for the path $\text{QD}_{\text{1D}}-\text{QD}_{\text{6E}}$, visiting 8 quantum dots over a round-trip distance of \SI{2.8}{\micro m} (see Figure~\ref{fig:figure3}d for the path). With the total round-trip fidelity extracted as $F_{\text{Round},\ket{\uparrow}}=99.4\pm0.1\%$, and some error coming from imperfect recovery gates, we retrieve an average lower bound of $F_{\text{BB}}=99.96\pm0.01\%$ per single bucket brigade transfer.  While basis states are initialised and recovered for this experiment, we note that during the shuttling any shuttling-induced rotation can bring the spin into a superposition state, making it susceptible to dephasing~\cite{vanriggelendoelman_2024_NatureCom_CoherentSpinQubit}. Moreover, if a recovery gate is imperfect, the state will be brought into a superposition and dephase as well, making it sensitive to small imperfections in the recovery gate~(see Supplementary Material~\ref{supp:recovery_gates}). Together these may explain the lower basis-state transfer fidelity in the other 8 round trip (Figure~\ref{fig:figure3}c). To study this effect, a superposition state is shuttled (Fig.~\ref{fig:figure3}b), resulting in a drop of coherence to $F_{\text{Round,}\ket{+}}=95.7\pm0.5\%$, indeed closer matching the typical basis-state transfer fidelities observed. The corresponding fidelity of $F_{\text{BB}\ket{+}}=99.68\pm0.04\%$ per hop is comparable to shuttling results in few-quantum dot systems~\cite{vanriggelendoelman_2024_NatureCom_CoherentSpinQubit,desmetHighfidelitySinglespinShuttling2025,noiriShuttlingbasedTwoqubitLogic2022}, though lower than measured for hopping qubits in double quantum dots \cite{Wang_2024_Science_OperatingSemiconductorQuantum}.

We populate ten spins in a sparse arrangement throughout the array, with the arrangement highlighted in Figure~\ref{fig:figure3}c, the densest arrangement allowing for individual control over neighbouring two-qubit interactions. We stress that this arrangement is optimised for regularity and qubit count, and not necessarily for qubit fidelity. With all qubits recovered into basis states, chevron maps are collected in two sets of five~(Fig.~\ref{fig:figure4}a) to allow for spin-resolved readout~(see Supplementary Material~\ref{supp:ExperimentalSequence} for the details on the readout). Even though five qubits are manipulated simultaneously, we crucially observe no noticeable crosstalk in the chevron experiments. The sparse occupation contributes to this by enhancing the frequency separation, given the finite spatial correlation of the g-factor~(Fig.~\ref{fig:figure1}e), and the local character of a microwave excitation~\cite{johnRobustLocalisedControl2025}. To quantify the crosstalk, the Rabi driving efficiency for different driving gates is studied. We observe an average suppression of $f_{\text{Rabi}}/f_{\text{Rabi,P}}\approx0.0085$ in the resonant crosstalk from driving at the closest qubit-site, compared to approximately $0.36$ for nearest-neighbour driving~(Fig.~\ref{fig:figure4}b-d). These results provide means for single-qubit addressability, without the strong reliance on either probabilistic or engineered~\cite{takakuraTripleQuantumDot2010,wangOptimalOperationPoints2021} frequency spacing for both electron and hole spins. Additionally, comparing the observed crosstalk to experiments with separate barrier gates~\cite{johnRobustLocalisedControl2025}, suggests that the merged barrier geometry lowers overall crosstalk.\\

To further study the effect of crosstalk we compare the Clifford Randomised Benchmarking (RB) fidelities on individually and simultaneously driven qubits in Figure~\ref{fig:figure4}e~\cite{emersonScalableNoiseEstimation2005,knillRandomizedBenchmarkingQuantum2008,magesanScalableRobustRandomized2011,gambettaCharacterizationAddressabilitySimultaneous2012,wuSimultaneousHighFidelitySingleQubit2025}. In both datasets, the native X\textsubscript{$\pi/2$} operations have equal time of $t_{\pi/2}=\SI{500}{ns}$ for all qubits. More details on the exact RB sequences can be found in the Methods section.
Single-qubit RB fidelities above the fault-tolerant surface code threshold of about 99\%~\cite{fowlerHighthresholdUniversalQuantum2009} are observed for all qubits, with generally no significant difference between the individual and simultaneous RB operation. As a result, this architecture is deemed promising for scaling to larger \AI{number} of qubits without experiencing a significant reduction in single-qubit control fidelity~\cite{wuSimultaneousHighFidelitySingleQubit2025}. \\
\begin{figure*}
    \centering
    \includegraphics[width=0.9\textwidth]{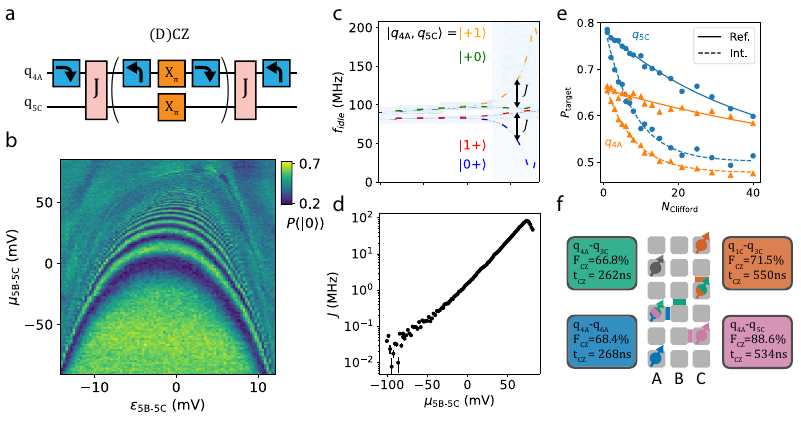}
    \caption{\textbf{Two Qubit Operation.} \textbf{a,} Schematic of a (decoupled) CZ operation between $q_{\text{4A}}$ and $q_{\text{5C}}$. The qubits are shuttled towards each other, at which point the exchange interaction $J$ is controlled and the qubits are separated. To decoupling single qubit dynamics the qubits can be shuttled back halfway for a decoupling pulse. \textbf{b,} Two-qubit exchange interaction is modulated through plunger gate voltages, resulting in the fingerprint plot like this one between the qubits at QD\textsubscript{5B} and QD\textsubscript{5C}. A decoupling sequence makes this fingerprint insensitive to g-factor variation. \textbf{c,} Fourier transforms of a superposition state evolution give the qubit spectrum, as a function of symmetric plunger gate voltage $\mu$ at $\varepsilon_{\text{5B-5C}}=0$. The results for different initial states (coloured) are overlapped, with the raw data available in Supplementary Material. The dashed lines give the fitted oscillation frequency. Indicated energy differences in the frequencies give the exchange interaction $J$. \textbf{d,} $J$ is plotted for different values of $\mu$.  \textbf{e,} CRB  for a reference (Ref.) and CZ-interleaved (Int.) sequence, gives a CZ fidelity of $F_{\text{CZ}}=88.6\pm1.6\%$. The displayed data is averaged over all elements of the character group for clarity. Raw data for the time-dependent oscillations and CRB can be found in Supplementary~\ref{supp:2Q}. \textbf{f,} An overview of the four calibrated qubit interactions, indicating the involved qubits, fidelity $F_{\text{CZ}}$ and total duration $t_{\text{CZ}}$. The location between the quantum dots involved in the interaction is coloured accordingly, as well as the involved qubits.} 
    \label{fig:figure5}
\end{figure*}

We proceed to connect several qubits to each other. When they are apart, no exchange interaction is measurable, so to enable two-qubit operations the desired spins are first bucket-brigade shuttled into proximity~(Fig.~\ref{fig:figure5}a). Once adjacent, two-qubit interaction is controlled through plunger-based exchange tuning~\cite{ivlevOperatingSemiconductorQubits2025a}. An example of the resulting fingerprint plot~\cite{reed_PRL_ReducedSensitivityChargeNoiseSymmetricGates} between qubits Q\textsubscript{4A} and Q\textsubscript{5C} is seen in Figure~\ref{fig:figure5}b, occupying QD\textsubscript{5B} and QD\textsubscript{5C}, respectively. From time-resolved oscillations along $\varepsilon_{5\text{B}-5\text{C}}=0$ detuning, we determine the exchange splitting for different initial states~(Fig.~\ref{fig:figure5}c-d), and see that the exchange interaction is controlled over three orders of magnitude, from $J_{\text{off}}\approx\SI{20}{kHz}$ to $J_{\text{on}}\approx\SI{80}{MHz}$. With this control over the exchange-interaction, we can calibrate a controlled-Z gate (CZ) between qubits Q\textsubscript{4A} and Q\textsubscript{5C}, resulting in a phase-shift of $\pi$ for the target qubit, conditioned on the state of the control qubit. The quality factor of these exchange oscillations and other details on two-qubit interactions can be found in Supplementary Material~\ref{supp:2Q}. \\

Finally, we use Character Randomised Benchmarking (CRB)~\cite{helsenNewClassEfficient2019,xueBenchmarkingGateFidelities2019} to benchmark the fidelity of this shuttling-mediated CZ gate at $F_{\text{CZ}}=88.6\pm1.6\%$ (see Figure~\ref{fig:figure5}e), placing it on par with shuttling-mediated CZ fidelities of devices with barrier gate control~\cite{noiriShuttlingbasedTwoqubitLogic2022,undsethWeightfourParityChecks2026}. We believe that this fidelity is mainly limited by single-qubit dephasing, given the relatively long gate duration of \SI{534}{ns} required for adiabatic shuttling and exchange coupling. In fact, we see significant improvements in the state coherence when decoupling pulses are interleaved~(Supplementary Material~\ref{supp:2Q}). Three additional qubit pairs were coupled~(Fig.~\ref{fig:figure5}f), yielding CZ fidelities around  70\%. These fidelities could potentially be improved through faster adiabatic shuttling, enhanced overall $T_2^*$ by in-plane magnetic field alignment~\cite{yuOptimisingGermaniumHole2026,hendrickxSweetspotOperationGermanium2024,mauro_2024_PRB_GeometryDephasingSweetSpot} and better control over the potential landscape. Moreover we expect that decreased variability of the g-tensor through material engineering~\cite{stehouwerGermaniumWafersStrained2023, stehouwerExploitingEpitaxialStrained2025} will allow for faster adiabatic shuttling and improved consistency of the CZ performance across the device.\\ 

Altogether we have shown that a merged barrier design enables us to go beyond quasi-1D systems and realise a 6$\times$6 quantum dot array. By probing the coherent evolution of shuttled spins, we are able to tune this array beyond the reach of the charge-sensors at the boundary. This opens up the possibility to study spin properties in a two-dimensional system at the spatial resolution of individual quantum dots, filling an existing gap in the device and metrology toolbox of spin-qubit systems. Furthermore, using two-way recovery gates we realise round-trip state-transfer fidelities up to $99.4\%$ for basis states and $95.7\%$ for superposition states over a distance of \SI{2.8}{\micro m}, suggesting that dedicated state-preparation and measurement zones may be attractive for large 2D arrays. Proceeding to initialise 10 spins in a sparse two-dimensional configuration, we show that crosstalk is reduced by a factor of 42 on average compared to dense equivalents, and achieve single qubit Clifford fidelities beyond the surface-code threshold across all ten qubits. Moreover using shuttling we bring together separated spins, and through plunger-only control of the exchange interaction realise two-qubit gates between these distant qubits.\\

This demonstrates key elements necessary to operate spin qubits in inherently 2D devices, \AI{mitigating operational overhead associated with one dimensional systems}. Moreover, with realistic improvements to coherence times~\cite{hendrickxSweetspotOperationGermanium2024,yuOptimisingGermaniumHole2026} and material improvements~\cite{stehouwerExploitingEpitaxialStrained2025}, we expect that beyond-threshold performance can also be realised for two qubit gates, ultimately over an extended distance. With reliable connections as a resource, distant links could be established, allowing for efficient codes relying on non-local connectivity~\cite{bravyi_2024_Nature_HighthresholdLowoverheadFaulttolerant}. Our results ultimately show that dedicated barrier gates may not be required, suggesting that a sparse, merged barrier processor can maintain local, high-fidelity single-qubit control while simultaneously providing size and connectivity advantages appealing to quantum simulation and fault-tolerant quantum computation alike. 

\section*{Methods}

\subsection*{Randomised benchmarking}
In this manuscript, several types of Randomised Benchmarking (RB) experiments are performed to quantify the fidelity of single and two qubit operations. To study the single qubit fidelity we perform the regular RB experiment~\cite{knillRandomizedBenchmarkingQuantum2008} on individual qubits, such that only a single qubit is operated on during one experimental run. Still all remaining qubits are loaded and read out during this experiment. To construct the Clifford operations used in the RB experiment, we use the generator set $\text{G}_{\text{X,Y}} = \{\text{X}_{\pi/2},\text{Y}_{\pi/2}\}$, as described in~\cite{lawrieSimultaneousSinglequbitDriving2023,wuSimultaneousHighFidelitySingleQubit2025} (see also Table~\ref{tab:Clifford_decomposition}). These native gates are implemented through EDSR driving, with the axis of rotation (either X or Y) being set by the phase of the applied microwave pulse. The amplitude of the microwave pulse is adjusted such that the native gate time for the $\pi/2$ rotation is \SI{500}{ns} for each qubit. The native gate times are synchronised to constitute a fair comparison to actual implementations, e.g. quantum error correction, which benefits from having comparable gate-time across the processor to avoid idle-induced decoherence. The Rabi frequency is dictated by the driving efficiency of the slowest qubit, while still ensuring high-quality driving across the qubits.\\

For the simultaneous RB, independent single-qubit Clifford sequences are executed on multiple qubits at the same time, following the method of~\cite{wuSimultaneousHighFidelitySingleQubit2025}. They are applied to five qubits at a time, utilising the remaining five qubits for spin-resolved readout. The experiment is then repeated for the remaining five qubits. All five sequences are started simultaneously, but since the Clifford operations generated by $\text{G}_{\text{XY}}$ decompose into a varying number of native gates, the total time may differ somewhat. Still, over the 200 Cliffords executed, this deviation averages out and the impact is negligible. This negligible effect is confirmed by extending the simultaneous RB sequence beyond \AI{saturation}, and observing similar fidelities (see Supplementary Material~\ref{supp:1Q}). 
\begin{figure}
    \centering
    \includegraphics[width=1\linewidth]{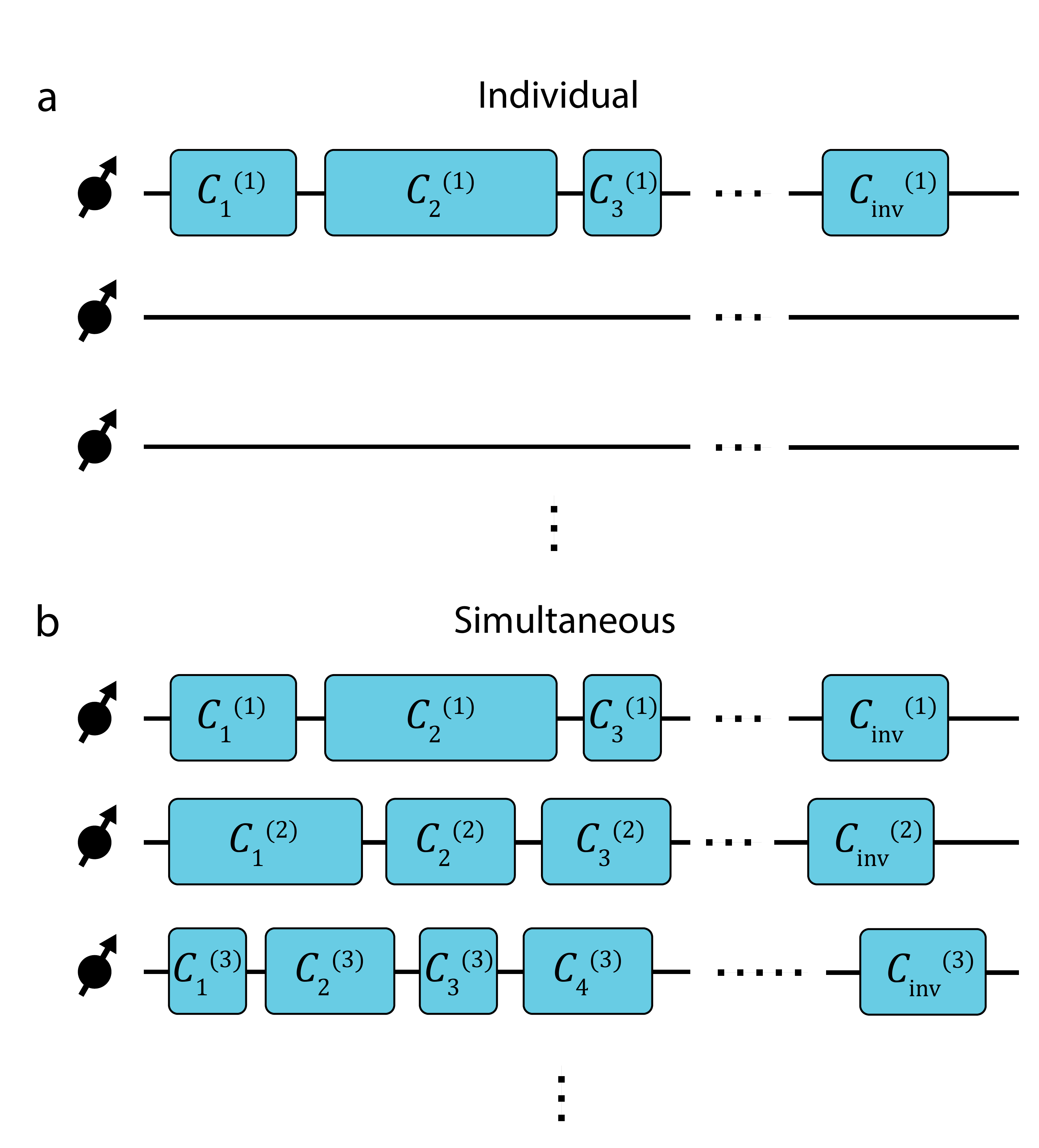}
    \caption{\textbf{RB experiment schematics.} \textbf{a,} Schematic of the individual RB experiment, in which only one qubit is subject to the randomised Clifford gates. \textbf{b,} Schematic of the simultaneous RB sequence in which multiple qubits are simultaneously subject to their own random Clifford sequence. Even though the native $\text{X}_{\pi/2}$ time is the same for all qubits, each Clifford of the used G\textsubscript{X,Y} decomposition consists of a varying amount of native gates (see Table~\ref{tab:Clifford_decomposition})}
    \label{fig:placeholder}
\end{figure}
The Character RB (CRB) sequence is used to study two-qubit gate fidelity~\cite{helsenNewClassEfficient2019}. By playing a reference sequence and then interleaving the two qubit gates, it is possible to extract the fidelity of just the two qubit gates from the difference of the observed decay curves. The major benefit of interleaving the CZ gate into a CRB sequence, compared to using regular interleaved RB~\cite{magesanEfficientMeasurementQuantum2012} is that the reference sequence can be composed of the single qubit Clifford set on both qubits, instead of the two qubit Clifford set~\cite{xueBenchmarkingGateFidelities2019}. Therefore, the reference sequence is generally of higher quality for CRB, facilitating the estimation of the CZ fidelity. \\

Since qubit decoherence is larger during idling than during driving, avoiding unnecessary idling is crucial when optimising for qubit coherence. To avoid idling between the interleaved two-qubit gates, a Clifford decomposition in which each single-qubit Clifford gate takes the same amount of time should be used. We use the G\textsubscript{X,Z} Clifford decomposition that is described in~\cite{xueBenchmarkingGateFidelities2019} (also see Table~\ref{tab:Clifford_decomposition}), in which all Clifford gate times are equal to that of two $\text{X}_{\pi/2}$ gates, at the expense of being more susceptible to low-frequency noise compared to the G\textsubscript{X,Y} decomposition. This results in the qubits being constantly driven between the implementation of the two qubit gates.
\begin{table}[]
\setlength{\extrarowheight}{1pt}
\caption{\textbf{Clifford decomposition overview. }Single-qubit Clifford decomposition used for the Randomised Benchmarking sequences. Since the Z-gates are implemented virtually, each Clifford in the G\textsubscript{X,Z} decomposition consists of two physical gates.}
\begin{tabular}{l|l|l}
\Xhline{1pt}
\hline\hline
Clifford                             & G\textsubscript{X,Y}                                                                          & G\textsubscript{X,Z}                               \\ \hline
C\textsubscript{1}  & [No Gate Executed]                                                                                                              & $\text{X}_{\pi/2}\text{X}_{-\pi/2}$                                 \\
C\textsubscript{2}  & $\text{X}_{\pi/2}\text{X}_{\pi/2}$                                                                                           & $\text{X}_{\pi/2}\text{X}_{\pi/2}$                                  \\
C\textsubscript{3}  & $\text{Y}_{\pi/2}\text{Y}_{\pi/2}$                                                                                           & $\text{Z}_{-\pi/2}\text{X}_{\pi/2}\text{X}_{\pi/2}\text{Z}_{\pi/2}$ \\
C\textsubscript{4}  & $\text{Y}_{\pi/2}\text{Y}_{\pi/2}\text{Y}_{\pi/2}\text{Y}_{\pi/2}$                                                                         & $\text{X}_{\pi/2}\text{Z}_{\pi/2}\text{Z}_{\pi/2}\text{X}_{\pi/2}$                              \\
C\textsubscript{5}  & $\text{X}_{\pi/2}\text{Y}_{\pi/2}$                                                                                           & $\text{X}_{\pi/2}\text{Z}_{-\pi/2}\text{X}_{\pi/2}\text{Z}_{\pi/2}$                             \\
C\textsubscript{6}  & $\text{X}_{\pi/2}\text{Y}_{\pi/2}\text{Y}_{\pi/2}\text{Y}_{\pi/2}$                                                                         & $\text{X}_{\pi/2}\text{Z}_{\pi/2}\text{X}_{\pi/2}\text{Z}_{-\pi/2}$                             \\
C\textsubscript{7}  & $\text{X}_{\pi/2}\text{X}_{\pi/2}\text{X}_{\pi/2}\text{Y}_{\pi/2}$                                                                         & $\text{X}_{-\pi/2}\text{Z}_{-\pi/2}\text{X}_{\pi/2}\text{Z}_{\pi/2}$                            \\
C\textsubscript{8}  & $\text{Y}_{\pi/2}\text{Y}_{\pi/2}\text{X}_{\pi/2}\text{Y}_{\pi/2}$                                                                         & $\text{X}_{-\pi/2}\text{Z}_{\pi/2}\text{X}_{\pi/2}\text{Z}_{-\pi/2}$                            \\
C\textsubscript{9}  & $\text{Y}_{\pi/2}\text{X}_{\pi/2}$                                                                                           & $\text{Z}_{-\pi/2}\text{X}_{\pi/2}\text{Z}_{\pi/2}\text{X}_{\pi/2}$                             \\
C\textsubscript{10} & $\text{Y}_{\pi/2}\text{X}_{\pi/2}\text{X}_{\pi/2}\text{X}_{\pi/2}$                                                                         & $\text{Z}_{-\pi/2}\text{X}_{\pi/2}\text{Z}_{\pi/2}\text{X}_{-\pi/2}$                            \\
C\textsubscript{11} & $\text{Y}_{\pi/2}\text{Y}_{\pi/2}\text{Y}_{\pi/2}\text{X}_{\pi/2}$ & $\text{Z}_{\pi/2}\text{X}_{\pi/2}\text{Z}_{-\pi/2}\text{X}_{\pi/2}$                             \\
C\textsubscript{12} & $\text{Y}_{\pi/2}\text{X}_{\pi/2}\text{Y}_{\pi/2}\text{Y}_{\pi/2}$ & $\text{Z}_{\pi/2}\text{X}_{\pi/2}\text{Z}_{-\pi/2}\text{X}_{-\pi/2}$                            \\
C\textsubscript{13} & $\text{X}_{\pi/2}$                                                                                                    & $\text{Z}_{-\pi/2}\text{X}_{\pi/2}\text{Z}_{\pi/2}\text{X}_{\pi/2}\text{Z}_{-\pi/2}$                   \\
C\textsubscript{14} & $\text{X}_{\pi/2}\text{X}_{\pi/2}\text{X}_{\pi/2}$                                                                                  & $\text{Z}_{\pi/2}\text{X}_{-\pi/2}\text{Z}_{-\pi/2}\text{X}_{-\pi/2}\text{Z}_{\pi/2}$                  \\
C\textsubscript{15} & $\text{Y}_{\pi/2}$                                                                                                    & $\text{X}_{\pi/2}\text{Z}_{\pi/2}\text{X}_{-\pi/2}$                                      \\
C\textsubscript{16} & $\text{Y}_{\pi/2}\text{Y}_{\pi/2}\text{Y}_{\pi/2}$                                                                                  & $\text{X}_{\pi/2}\text{Z}_{-\pi/2}\text{X}_{\pi/2}$                                      \\
C\textsubscript{17} & $\text{Y}_{\pi/2}\text{X}_{\pi/2}\text{Y}_{\pi/2}\text{Y}_{\pi/2}\text{Y}_{\pi/2}$                                                                & $\text{X}_{-\pi/2}\text{Z}_{\pi/2}\text{Z}_{\pi/2}\text{X}_{-\pi/2}\text{Z}_{-\pi/2}$                  \\
C\textsubscript{18} & $\text{Y}_{\pi/2}\text{Y}_{\pi/2}\text{Y}_{\pi/2}\text{X}_{\pi/2}\text{Y}_{\pi/2}$                                                                & $\text{X}_{-\pi/2}\text{Z}_{-\pi/2}\text{Z}_{-\pi/2}\text{X}_{-\pi/2}\text{Z}_{\pi/2}$                 \\
C\textsubscript{19} & $\text{X}_{\pi/2}\text{X}_{\pi/2}\text{Y}_{\pi/2}$                                                                                  & $\text{X}_{\pi/2}\text{Z}_{-\pi/2}\text{X}_{\pi/2}$                                      \\
C\textsubscript{20} & $\text{Y}_{\pi/2}\text{X}_{\pi/2}\text{X}_{\pi/2}$                                                                                  & $\text{X}_{\pi/2}\text{Z}_{\pi/2}\text{X}_{\pi/2}$                                       \\
C\textsubscript{21} & $\text{Y}_{\pi/2}\text{Y}_{\pi/2}\text{X}_{\pi/2}$                                                                                  & $\text{Z}_{-\pi/2}\text{X}_{\pi/2}\text{Z}_{\pi/2}\text{X}_{\pi/2}\text{Z}_{\pi/2}$                    \\
C\textsubscript{22} & $\text{X}_{\pi/2}\text{Y}_{\pi/2}\text{Y}_{\pi/2}$                                                                                  & $\text{Z}_{-\pi/2}\text{X}_{\pi/2}\text{Z}_{\pi/2}\text{X}_{-\pi/2}\text{Z}_{-\pi/2}$                  \\
C\textsubscript{23} & $\text{Y}_{\pi/2}\text{X}_{\pi/2}\text{Y}_{\pi/2}$                                                                                  & $\text{X}_{\pi/2}\text{X}_{\pi/2}\text{Z}_{\pi/2}$                                       \\
C\textsubscript{24} & $\text{Y}_{\pi/2}\text{X}_{\pi/2}\text{X}_{\pi/2}\text{X}_{\pi/2}\text{Y}_{\pi/2}$                                                                & $\text{X}_{-\pi/2}\text{X}_{-\pi/2}\text{Z}_{-\pi/2}$  \\

\hline\hline
\Xhline{1pt}

\end{tabular}
\label{tab:Clifford_decomposition}
\end{table}
\section*{Acknowledgements}
We thank Corentin D\'eprez and Valentin John for in-depth discussions on the device design. We are grateful for the continued software support provided by Sander de Snoo and Daan Bijl across many levels of the experimental stack. We also thank the technical staff, in particular Olaf Benningshof, Tom Orton, Jason Mensingh, Jelle Haanstra, Raymond Vermeulen and Raymond Schouten for repairing and maintaining the cryogenic, vacuum and electronic aspects of the experiment throughout the two years of measurements. 
\section*{Data Availability}
The analysis code and raw data supporting the findings of this study will be openly available in the Zenodo repository \href{www.doi.org/10.5281/zenodo.23102383}{10.5281/zenodo.2310238} at the time of publication~\cite{zenodo_Data}.\\
\section*{Funding}
We acknowledge support by the Dutch Research Council through an NWO ENW grant (M20.378) and by the Dutch National Growth Fund (NGF), as part of the Quantum Delta NL programme. We acknowledge the IGNITE project of the European Union’s Horizon Europe Framework Programme under grant agreement No. 101069515. This research was sponsored in part by the Army Research Office (ARO) under Award No. W911NF-23-1-0110. This research was sponsored in part by The Netherlands
Ministry of Defence under Awards No. QuBits R23/009. The views, conclusions, and recommendations contained in this document are those of the authors and are not necessarily endorsed nor should they be interpreted as representing the official policies, either expressed or implied, of the Army Research Office (ARO) or the U.S. Government, and also not the Netherlands Ministry of Defence. The U.S. Government and the Netherlands Ministry of Defence is authorized to reproduce and distribute reprints for Government purposes, notwithstanding any copyright notation herein.

\section*{Competing Interest}
A.S.I. is an inventor on a patent application on the merged barrier architecture
filed by Delft University of Technology under the application number NL38924. M.V. and G.S. are founding advisors of Groove Quantum BV, with M.V. and G.S. declaring equity interests. The remaining authors declare that they have no competing interests.

\section*{Supplementary Material}
Additional information on the device, experimental procedure and results can be found in the Supplementary Material~\cite{supp_reference}.


\clearpage
\input{main_bbl.tex}
\clearpage

\input{supp_content}
\end{document}

%% file: main_bbl.tex
%

%% file: supp_content.tex
\onecolumngrid

\title{Supplementary Material: \textbf{Sparse qubit operation in a 6$\times$6 quantum dot array}} 


\author{Alexander S. Ivlev}
\thanks{}
\affiliation{QuTech and Kavli Institute of Nanoscience, Delft University of Technology, PO Box 5046, 2600 GA Delft, The Netherlands}
\author{Harold L. Co}
\affiliation{QuTech and Kavli Institute of Nanoscience, Delft University of Technology, PO Box 5046, 2600 GA Delft, The Netherlands}
\author{Damien R. Crielaard}
\affiliation{QuTech and Kavli Institute of Nanoscience, Delft University of Technology, PO Box 5046, 2600 GA Delft, The Netherlands}
\author{Alice Petrillo}
\affiliation{QuTech and Kavli Institute of Nanoscience, Delft University of Technology, PO Box 5046, 2600 GA Delft, The Netherlands}
\author{Setareh Kazemzadeh}
\affiliation{QuTech and Kavli Institute of Nanoscience, Delft University of Technology, PO Box 5046, 2600 GA Delft, The Netherlands}
\author{Giordano Scappucci}
\affiliation{QuTech and Kavli Institute of Nanoscience, Delft University of Technology, PO Box 5046, 2600 GA Delft, The Netherlands}
\author{Menno Veldhorst}
\affiliation{QuTech and Kavli Institute of Nanoscience, Delft University of Technology, PO Box 5046, 2600 GA Delft, The Netherlands}

\maketitle

\renewcommand\thefigure{S\arabic{figure}}
\setcounter{figure}{0}

\section{Device And Fabrication}
\label{supp:fabrication}
The device discussed in this manuscript, as well as earlier device generations, are fabricated on a strained Ge/SiGe heterostructure on top of a Si(001) wafer~\cite{lodari_MatQuantTech_LowPercolationDensity2021} \AI{grown by} reduced-pressure chemical vapour deposition on top of a  Si\textsubscript{0.2}Ge\textsubscript{0.8} strain relaxed buffer, the heterostructure comprises a 16 nm strained Ge quantum well,  a \SI{55}{nm} Si\textsubscript{0.2}Ge\textsubscript{0.8} spacer, and a native SiO\textsubscript{x} layer. \\

\begin{figure}
    \centering
    \includegraphics[width=0.6\linewidth]{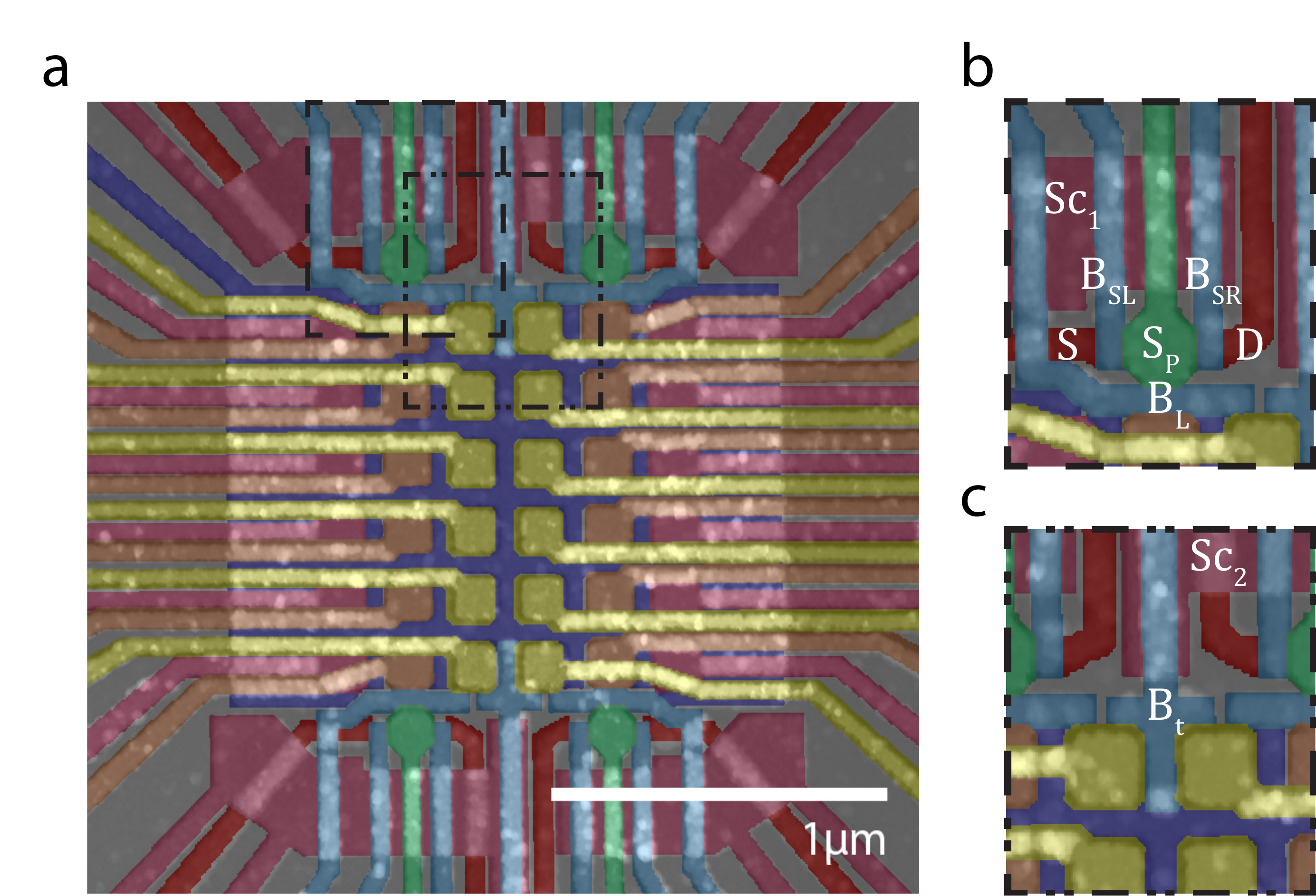}
    \caption{\textbf{Device AFM.} \textbf{a,} The entire nanostructure of the device is depicted, with false colour applied to all gates. Specific regions in \textbf{b} and \textbf{c} are highlighted with dashed lines, specifying different gates described in the text. Due to the device symmetry, and the labelling of other gates in the main, only the top-left SET and the additional top barrier gate B\textsubscript{t} are highlighted to avoid clutter.}
    \label{fig:supp_deviceDetails}
\end{figure}

On top of the heterostructure, the gates-stack consist of five metal layers. The first metallic layer is platinum (\SI{30}{nm}), which diffuses into the heterostructure to provide the ohmic source (S) and drain (D) contacts, used for radio-frequency reflectometry. After the ohmics, a \SI{7}{nm} layer of AlOx is grown through atomic layer deposition (ALD) at 300\degree C. The other ALD layers separating the gates are \SI{5}{nm}. The subsequent four Ti/Pd layers mainly consist of the merged barrier (B\textsubscript{m}) (3/17\,\SI{ }{nm}), followed by three plunger gate layers (3/27\,\SI{ }{nm}, 3/27\,\SI{ }{nm} and 3/37\,\SI{ }{nm} respectively). The plunger gates of the same colour in main Figure~\ref{fig:figure1} are deposited in the same layer and have a gate to gate pitch of \SI{200}{nm}. The B\textsubscript{m} in addition screens the heterostructure from the plunger gate fanout to prevent unwanted accumulation. The other screening gates Sc\textsubscript{i} deplete the fanout of the single hole transistors and are deposited in the same layer as B\textsubscript{m}. The plunger gates (P\textsubscript{S,i}) of the different single hole transistors (SHT) are deposited in the same layer as the second column of array-plunger gates, with the barrier gates around it being deposited together with the first plunger gate column. These SHT barrier gates consist of a left and right barrier gate B\textsubscript{SL/R} that tune the tunnel coupling to the ohmic leads, and a loading barrier B\textsubscript{L} controlling the loading into the array. With the exception of the direct charge sensing data in Supplementary Material~\ref{fig:incoherent_dots} only the top-left charge sensor is used during the entirety of the experiment, with the remaining sensors being depleted. This top-left sensor also functions as the reservoir to load and unload charges into the array. Additional barrier gates B\textsubscript{t/b} are patterned in the same layer as the SHT barrier gates, and allow one to control the merged barrier B\textsubscript{m} voltage without critically impacting tunnel coupling between QD\textsubscript{1C}-QD\textsubscript{1D} and QD\textsubscript{6C}-QD\textsubscript{6D}. This simplified the calibration of Pauli Spin Blockade (PSB)~\cite{onoCurrentRectificationPauli2002}, speeding up initial screening of the device, but during the presented experiment these gates are kept at a constant voltage, comparable to the voltage of B\textsubscript{m}. The gates discussed here are labelled in more detail in Figure~\ref{fig:supp_deviceDetails}.

\section{Earlier Devices}
\label{supp:EarlierDevices}
The device presented in the main section of the article is the third generation of the merged barrier device family that was attempted for qubit measurements. In this section we want to provide learnings from these earlier versions, to give insight in choices made in design and tuning and the device yield. \\

\begin{figure}
    \centering
    \includegraphics[width=0.8\linewidth]{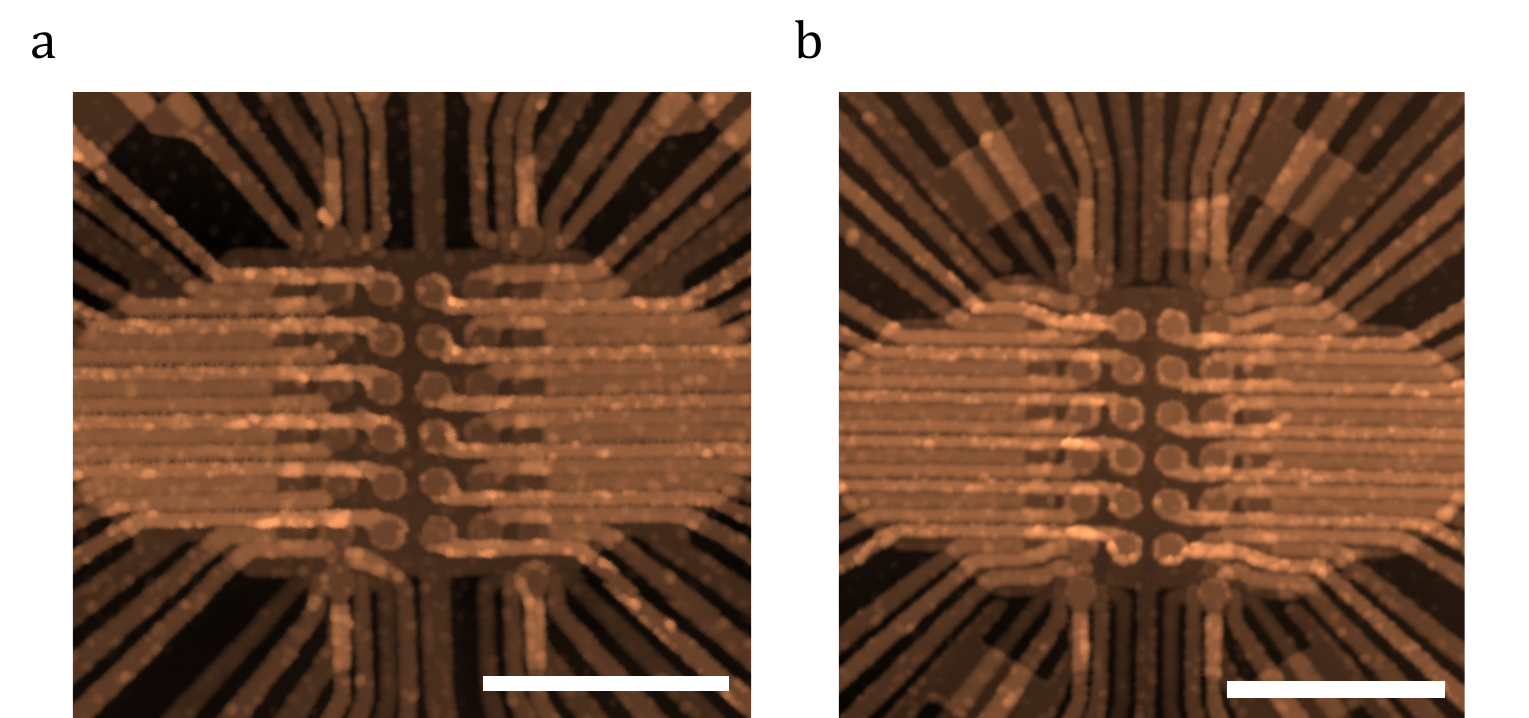}
    \caption{\textbf{Earlier Devices.} \textbf{a,} and \textbf{b,} display Atomic Force Microscope (AFM) images of respectively the first and second iteration of the $6\times6$ merged barrier design. The white scalebars at the bottom indicate \SI{1}{\micro m}.}
    \label{fig:old_device}
\end{figure}

The first version of the design~(Figure~\ref{fig:old_device}\textbf{a}) generally had similar dimensions as earlier Ge/SiGe devices~\cite{johnRobustLocalisedControl2025}, with the exception of the merged barrier gate. As a further departure from earlier design, not all ohmic gates were patterned up to the Single Hole Transistors (SHTs) but instead accumulation gates were tried, as is typical in Si/SiGe devices~\cite{philips_2022_Nature_UniversalControlSixqubit,unseldBasebandControlSingleelectron2024}. The goal of this ohmic design-change, was to attempt to improve the Coulomb peaks of the SHT. While these accumulation gates in principle produced working sensors, we did not manage to tune them to a noticeably higher quality than with locally annealed ohmic gates, and this design variation was therefore abandoned in the subsequent versions. Still, to our knowledge this is the first realisation of accumulation gates in Ge/SiGe, and shows that the principles of using them are generally sound. Perhaps further design iterations would lead to enhanced charge sensor performance. \\

While in this first design variation sensors and quantum dots could be formed, we noticed that the charge-sensor signal was highly dependent on the DC voltages applied on the plunger gates of the array, making tune-up of the entire array not possible. We understand this to result from 2D hole gas (2DHG) accumulation underneath the fanout of the array, to which the reflectometry signal would couple strongly, despite the separation of hundreds of nanometres.\\

To mitigate the impact of the array voltages on the sensor signal, the screening gates have been extended starting from design version 2~(Figure~\ref{fig:old_device}\textbf{b}) to prevent spurious coupling to a 2DHG. Moreover, additional barriers were introduced to better control charge loading into the array. This allowed us to utilise all four charge sensors and attempt direct charge-sensing throughout the array. Finally, more distant charge sensors were moved towards the inner column to increase the sensitive area. Despite this, sensing the inner two rows remained difficult to do reliably. \\

\begin{figure}
    \centering
    \includegraphics[width=\linewidth]{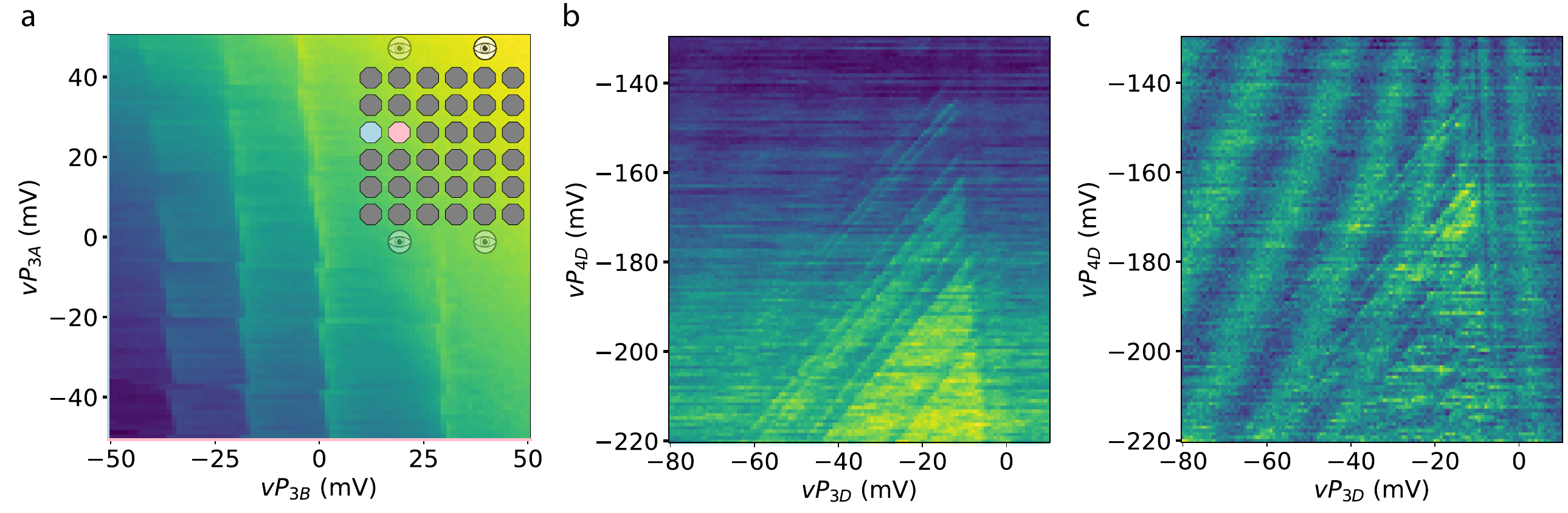}
    \caption{\textbf{Readout Features}. \textbf{a,} Example of cascade readout, where QD\textsubscript{3A} (lightblue) and QD\textsubscript{3B} (pink) were sensed with the top right charge-sensor. \textbf{b,} Charge loss during the shuttling can be picked up by a charge-sensor when the state is nominally back in the (1,1) occupation of the readout pair. The exact charge-loss features depend on the origin of the charge loss, here the features run along the (extended) interdot transitions, since presumably the charge got lost after tunneling from QD\textsubscript{3D} to QD\textsubscript{4D}. Subfigure \textbf{c,} corresponds to the same scan as in \textbf{b,} but now resolving coherent phase information \AI{using spin-to-charge conversion through PSB}, demonstrating how charge-loss may look in an SED.}
    \label{fig:readout_features}
\end{figure}

In general, we observe that the sensitivity of direct charge-sensing degrades when the array is sparsely populated. This stands in contrast to the typically held belief that the presence of in-between charges screens the detection of further-lying charges, and instead it seems that the opposite is true. In our attempts to extend charge-sensing we implemented cascade readout~\cite{vandiepenElectronCascadeDistant2021,chittock-woodRadiofrequencyCascadeReadout2026,nicklEightqubitOperation3002026}, ultimately achieving a sensing distance of 4 quantum dots horizontally and 3 vertically (see Figure~\ref{fig:readout_features}). With a gate-pitch of \SI{200}{nm}, this corresponds to a straight-line distance of \SI{1}{\micro m}, far beyond the maximal charge sensing distance of up to $\approx$\SI{500}{nm} that we typically observe. This shows for the first time, that cascade charge-readout beyond the next nearest PSB pair is achievable. Still, this cascade-based tuning is very susceptible to drift, and moreover requires one to park quantum dots close to interdot transitions during readout. This makes it difficult to tune to a specific charge-state in a large array, particularly if that is a sparse-charge occupation. \\

Given the challenges in charge sensing, some charge stability diagrams of the middle rows remained unconvincing, and the qubit-based probing through SEDs as described in the main text was developed. Here we discovered that not all quantum dots could be reached coherently, and in particular that the charge can be lost during the shuttling process (see Figure~\ref{fig:readout_features} for an example). In this device only 6 quantum dots could be reached from the initial QD\textsubscript{1C}-QD\textsubscript{1D} pair, with the remaining paths being cut off due to loss in coherence. We attributed this to too low tunnel coupling, which we could not increase further with the barrier gate, as other spurious quantum dots would form. Plunger-based tuning also was not sufficient to within the range of our electronics.\\

Hence, for the final version of the merged-barrier design, as reported in this manuscript, the dimensions of the different gates have been adjusted, with the hope of increasing the width of the wavefunction overlap between neighbouring quantum dots and thus enhance tunnel coupling. In particular, the regular octagons that typically are used in the designs of our quantum chips have been squared off slightly. While this may still not be the optimal design, it seems to produce homogeneous enough tunnel coupling to allow shuttling throughout most of the array. In this final version, additional barrier gates have been introduced into the array to allow for individual tuning of the initialisation and readout of the qubits, without affecting the merged barrier or plunger gates. We note however, that this was simply for convenience and flexibility, as the previous version of the devices made and shuttled spin qubits without this. 

\section{Experimental Setup}
\label{supp:experimentalSetup}
The device operates in a Bluefors LD-400 dilution refrigerator at a nominal temperature of \SI{25}{mK}. A uniaxial \SI{3}{T} solenoid magnet from American Magnetics is mounted on the 4 Kelvin stage to provide the \SI{25}{mT} magnetic field nominally parallel with the sample plane. The magnet is operated in driven mode during the entirety of the experiment. \\

DC bias is provided using in-house voltage-sources, and combined with arbitrary waveform generator (AWG) pulses using bias-tees on an in-house PCB on which the quantum chip is glued. The AWGs used are Keysight M3202A, whose signals were attenuated between 22dB and 9dB depending on the range required for voltage pulsing. The readout excitation signal was also produced with the M3202A, and undergoes with 85dB attenutation before reaching the in-house NBTiN resonators. Readout was further enabled through a Cosmic Microwave CITLF3 low-noise amplifier with 33dB nominal gain, paired with an in-house room-temperature amplifier and a Keysight M3102A digitiser. 

\section{Spin Evolution Diagram Features}
\label{supp:SED}
\begin{figure}
    \centering
    \includegraphics[width=0.8\linewidth]{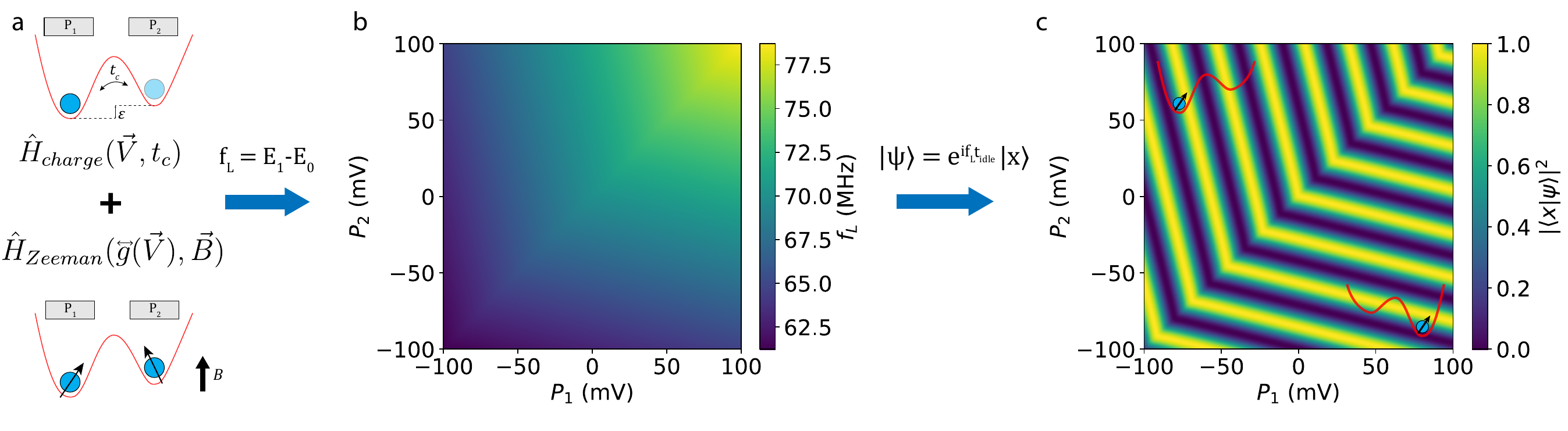}
    \caption{\textbf{Spin Evolution Diagram Simulation.} \textbf{a,} The Hamiltonian used to simulate the quantum dot system consists of a charge part $\hat{H}_{charge}$, responsible for the charge transport and the coupling between the quantum dots, and a spin part $\hat{H}_{Zeeman}$ that lifts the spin-degeneracy under an applied magnetic field. The plunger gate voltages $(P_1,P_2)\equiv \vec{V}$ impact the eigenstates both through the charge-distribution and the voltage-dependence of the g-factor. \textbf{b,} The local Larmor frequency is computed as the energy splitting of the lowest two Hamiltonian eigenstates, as function of the plunger gate voltages, for some simplified configuration. \textbf{c,} The phase-pickup of a superposition state for a constant idling time $t_{\text{idle}}$ can be computed throughout the voltage space, producing a Spin Evolution Diagram. Due to the different g-factor susceptibility in the two quantum dots under P\textsubscript{1} and P\textsubscript{2}, two regions separated by an interdot transition are distinguishable. For this simulation $g=0.2$ for both quantum dots at $\vec{V}=0$, and their quantisation axes are aligned. $t_c=\SI{50}{GHz}$ is used.}
    \label{fig:SED_simulation_steps}
\end{figure}

In the main manuscript we introduced the method of Spin Evolution Diagrams (SEDs) to tune the charge occupation of quantum dots outside the range of a charge sensor. This method is based on the coherent Larmor precession of a spin state in superposition. Since the spin evolution depends on the Larmor frequency, which itself typically dependents on the quantum dot site which the spin occupies and the electrostatic potential, it is possible to deduce which site the spin is located at~\ref{fig:SED_simulation_steps}. However, since the Larmor frequency can also vary within a particular quantum dot, the features are not always trivial to interpret. This section is meant to provide an overview of the different features and regimes that are typically observed during SED-based tuning, and provide a guide for future experiments to interpret the behaviour observed in the system. \\
In the simulations discussed in this section, it is assumed that the magnetic field $|B|=\SI{25}{mT}$ and the idling duration is $t_{\text{idle}}=\SI{500}{ns}$, in order to better mimic the results in the manuscript. Moreover, for simplicity the simulation only includes idling at a given location in voltage space, and no phase-pickup or other (in)coherent transitions before and after the idling time are accounted for. The code used for the simulations \AI{is available through Zenodo~\cite{zenodo_Data}}. 
\subsection{G-factor Variation}
\begin{figure}
    \centering
    \includegraphics[width=0.6\linewidth]{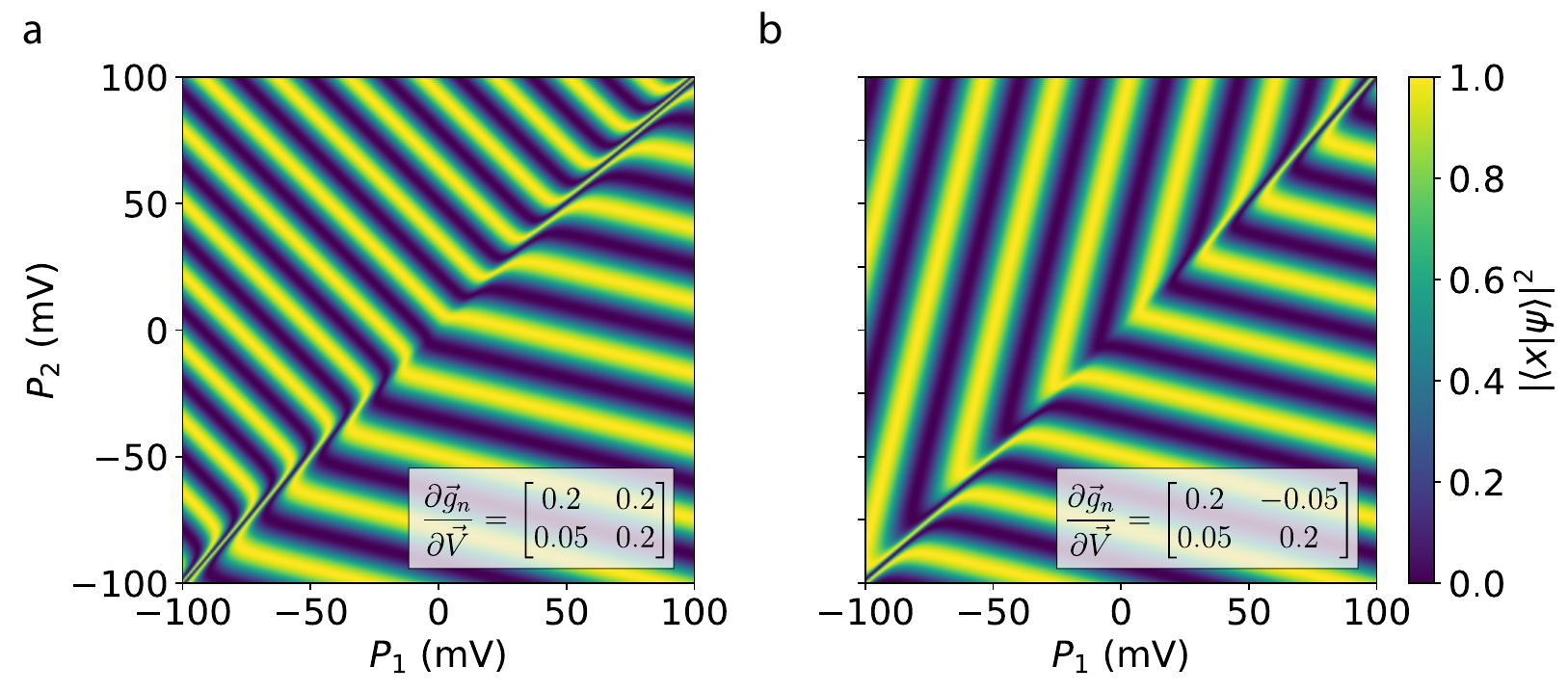}
    \caption{\textbf{Impact of g-factor sensitivity.} \textbf{a-b,} Changing the g-factor sensitivity to the voltages impacts the spacing and slope of the phase-fringes.}
    \label{fig:SED_susceptibility}
\end{figure}

The SED presented in Figure~\ref{fig:SED_simulation_steps}c illustrates a simple example, \AI{in which} the g-tensors \AI{are} aligned, and the effective g-factors of both quantum dots are equal $ \left. g_1=g_2\right|_{\vec{V}=0}=0.2$. The different g-factor susceptibility still allows one to recognise the interdot transition, with $\partial g_1/\partial \vec{V}=(0.2,0.05)/\SI{}{mV}$ and $\partial g_2/\partial \vec{V}=(0.05,0.2)/\SI{}{mV}$. For compact notation, we'll combine the g-factors in a single vector $(g_1,g_2)=\vec{g}_n$, with the subscript reminding us that this is the effective g-factor, and not a g-tensor. With the material variation of the germanium devices, the case of Figure~\ref{fig:SED_simulation_steps}c is a simplified one, however, and several aspects can be studied to better appreciate the zoo of SEDs that can be observed in typical devices. Firstly, the g-factor susceptibility $\partial \vec{g}_n/\partial \vec{V}$ need not be a symmetric matrix. For example, we can change the voltage dependence of $g_1$, to make the side-plunger voltage $P_2$ have the same effect as $P_1$~(see Figure~\ref{fig:SED_susceptibility}). Alternatively, the sign of the side-plunger susceptibility can fully change, something that has been consistently observed experimentally in our device and in previous studies~\cite{johnRobustLocalisedControl2025}.\\

\begin{figure}
    \centering
    \includegraphics[width=0.65\linewidth]{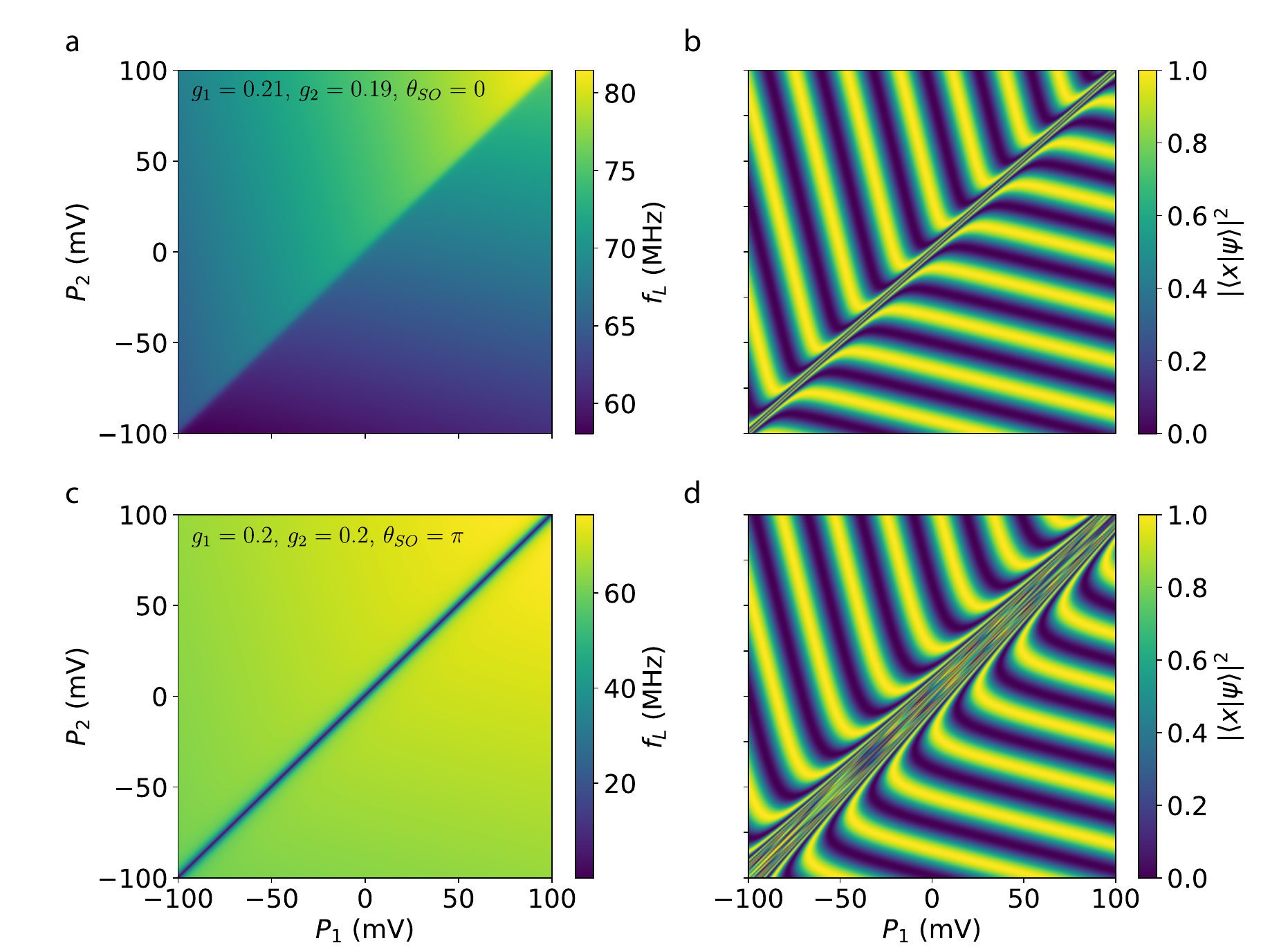}
    \caption{\textbf{Impact of g-tensor variation.} \textbf{a-b,} Changing the g-factor gives a discontinuity at the interdot transition, regardless of phase-fringes. \textbf{c-d,} A spin-orbit angle creates a drop change in Larmor frequency around the interdot transition, resulting in a clearly visible anticrossing.}
    \label{fig:SED_gfacot_variation}
\end{figure}

Without altering the g-factor susceptibility, the g-tensor itself can also change. The simplest parameter to change is the magnitude of the g-factor, and in Figure~\ref{fig:SED_gfacot_variation}a-b one finds the situation where $\left. \vec{g}_n\right|_{\vec{V}=0}=(0.21,0.19)$. The sudden change in g-factors between the quantum dots created a clear interdot transition. By changing the direction of the quantisation axes one can also create a strong apparent interdot transition, since the Larmor frequency decreases strongly in the region where the spin is delocalised. In Figure~\ref{fig:SED_gfacot_variation}c-d this effect is maximally visible due to the complete anti-alignment of the quantisation axes by an angle of $\theta_{SO}=\pi\,$rad.

\subsection{Dot-dot coupling}
\begin{figure}
    \centering
    \includegraphics[width=0.6\linewidth]{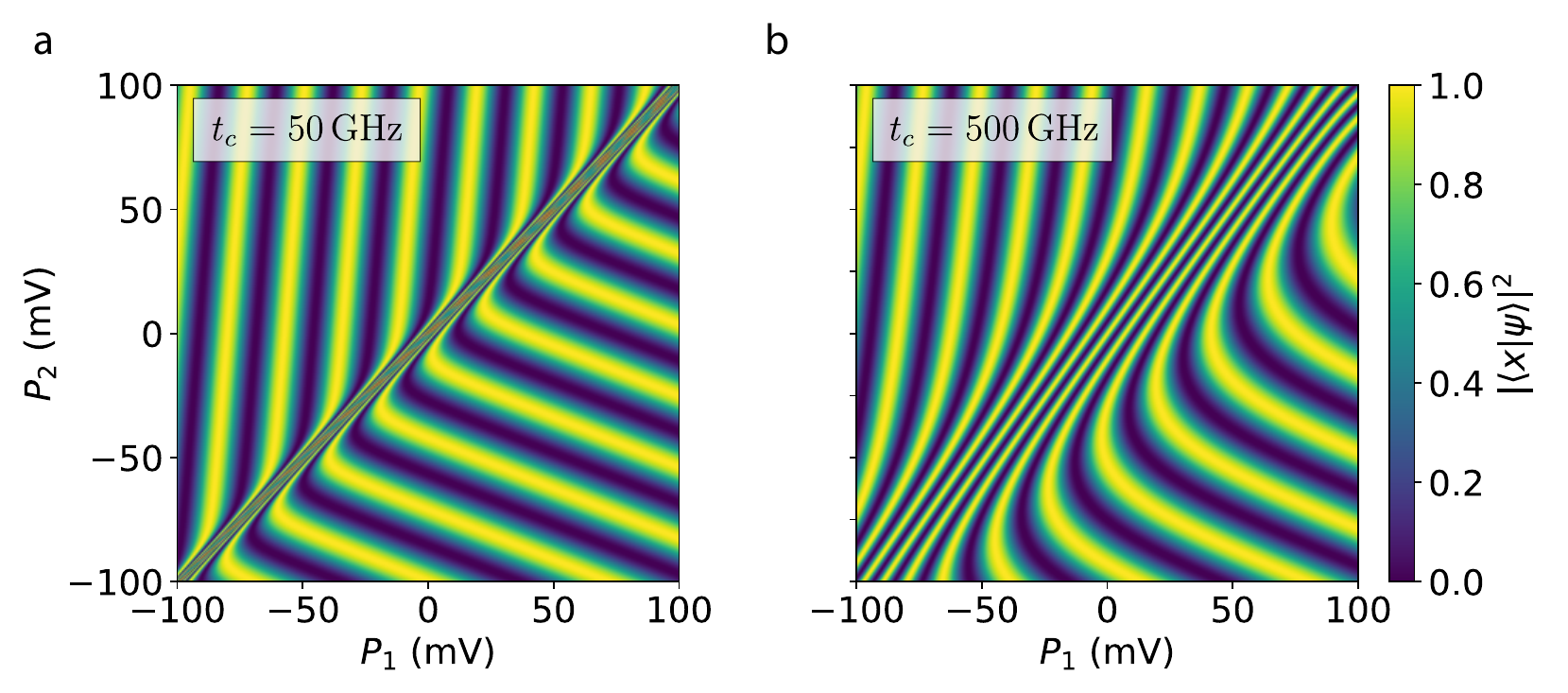}
    \caption{\textbf{Impact of tunnel coupling.} As the tunnel coupling increases from $t_c=\SI{50}{GHz}$ in \textbf{a} to $t_c=\SI{500}{GHz}$ in \textbf{b}, the exact location of the interdot becomes less discernable.}
    \label{fig:SED_tc_effect}
\end{figure}

Most material variations of the spin-Hamiltonian create clear interdot transitions, making SED-based tuning straightforward. In Figure~\ref{fig:SED_tc_effect}a, a double quantum dot system is simulated with realistic variation parameters across the two sites, making the interdot transition clearly visible. Variation in tunnel coupling can however wash out the dot-dot Zeeman variation, by hybridising the charge states of the different quantum dot sites. As we increase the tunnel coupling by an order of magnitude to $t_c=\SI{500}{GHz}$, we observe that the interdot transition is less visible Figure~\ref{fig:SED_tc_effect}b. While this tunnel coupling is large, it is within a reasonable parameter space, and in practice we do see signatures of strong hybridisation in the SEDs. \\

\begin{figure}
    \centering
    \includegraphics[width=0.9\linewidth]{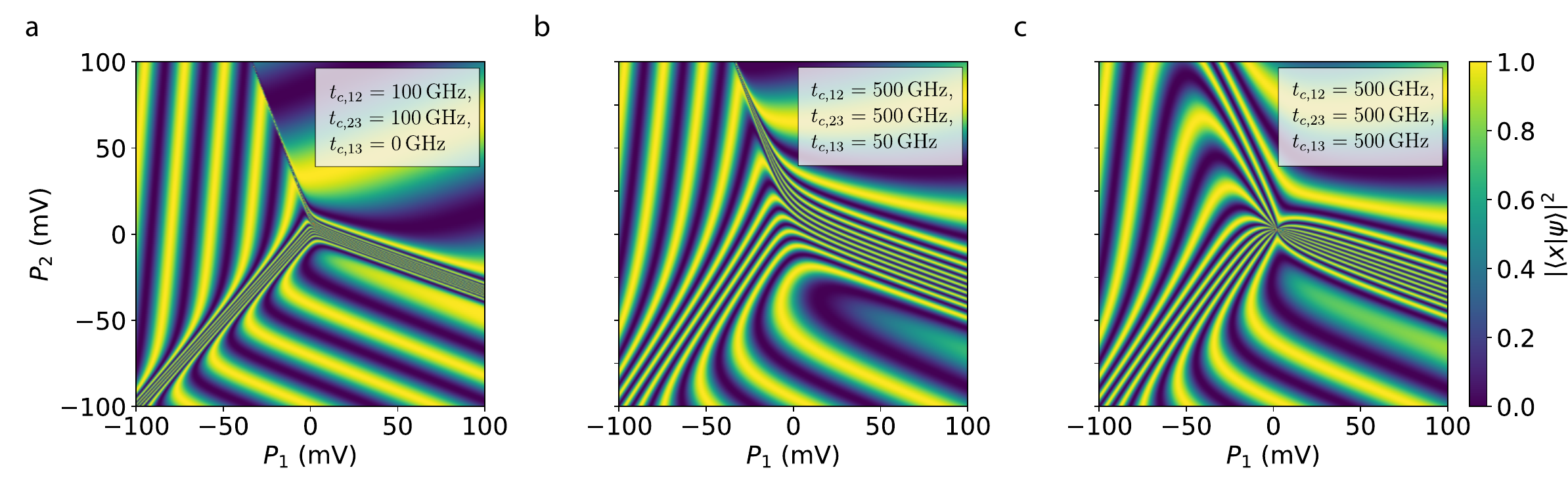}
    \caption{\textbf{Coupling to third quantum dot.} \textbf{a,} A third quantum dot is introduced, which doesn't couple to QD\textsubscript{1} ($t_{c,13}=0$), mimicking a linear chain. \textbf{b-c,} Introducing more coupling between the quantum dots creates highly non-linear effects that are also observed in the experimental data.}
    \label{fig:SED_tripledot}
\end{figure}

The double quantum dot system in which spin-transport is being studied could also be coupled to another quantum dot. To keep the analysis simple, we introduce a single third quantum dot as this captures the general behaviour and is typically sufficient to describe the SEDs that we observe during tuning. This third quantum dot can either be fully decoupled from the first quantum dot~(Figure~\ref{fig:SED_tripledot}a), effectively forming a linear chain. Often, a finite but asymmetric coupling is more reasonable~(see Figure~\ref{fig:SED_tripledot}b), and genuine triple dot system behaviour can not be fully ruled out either~(see Figure~\ref{fig:SED_tripledot}c) when interpreting these diagrams. The highly non-linear effects of the phase-fringes observed in this simulation qualitatively resemble experimental data.\\

Clearly, once more quantum dots are involved in the system, in particular when the tunnel coupling is large, the transition may be more challenging to discern. Moreover, it should be noted that this is a simplified study, and in a general situation there may be a higher order voltage dependence of the g-factor and the g-tensor more generally. Based on the plunger-based exchange tunability, we furthermore expect that the tunnel-coupling is voltage dependent, something that is not captured in the current simulation. 

\subsection{Path-Dependant Effects}
In the presented SED model, the observed fringes at any given voltage point are a result from the phase pickup due to the Larmor frequency at that given point. In actual SED experiments, there is some phase-pickup during the shuttling towards a particular voltage point as well. Any common phase-pickup during most of the shuttling is irrelevant here, but the phase-pickup when going to the particular point in the SED map may impact the map. Generally, this effect is small because the ramp time of \SI{20}{ns} is short compared to the idling time of \SI{500}{ns} at each voltage point.\\

Next to this phase-pickup, the shuttling path may cross an interdot transition resulting in diabatic spin transition. Since the degree of diabaticity depends on the ramp rate and the hopping gate executed depends on the time spent in a quantum dot~\cite{Wang_2024_Science_OperatingSemiconductorQuantum}, there may be additional hopping induced fringes observed in the SEDs. While these have been observed during the initial calibration, they did not substantially impact the tuning effort. \\
\subsection{Incoherent Transitions}
Finally, we note that there are some incoherent effects that also play a role during the collection of SEDs. For one, a charge may be lost during shuttling, resulting in an artefact within a particular region. These artefacts can be clearly distinguished by the charge-sensor from decoherence, as the total amount of charges during charge-readout is different than expected after spin-to-charge conversion. At the current voltage configuration these charge-loss artifacts are particularly prevalent in the bottom right of the device~(see main Figure~\ref{fig:figure1}). Spin-decoherence is also possible, for example due to a low $T_2^*$ at a particular location, but potentially also due to a low-fidelity bucket-brigade transition. We typically observe such low fidelity transitions quantum dot pairs that seem to have a small $t_c$, suggesting that operating at high $t_c$ would be beneficial in future work. 

\section{Incoherent Quantum Dot Sites}
\label{supp:incoherent_sites}
\begin{figure}
    \centering
    \includegraphics[width=\linewidth]{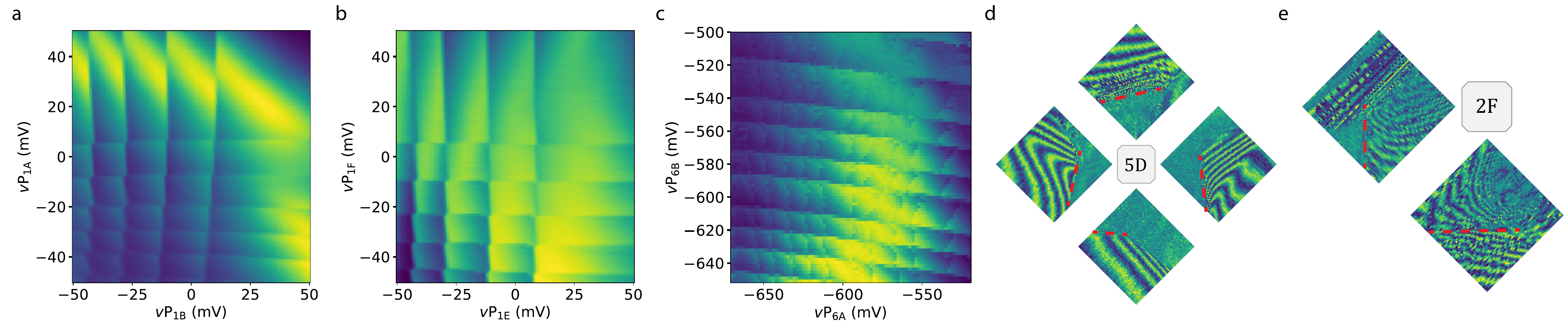}
    \caption{\textbf{Sensing Inaccessible Quantum Dots}. \textbf{a-c,} Charge Stability Diagrams obtained through direct charge sensing, showing the ability to form quantum dots QD\textsubscript{1A} (\textbf{a}), QD\textsubscript{1E} and QD\textsubscript{1F} (\textbf{b}),  and QD\textsubscript{2B} (\textbf{c}). \textbf{d,} Spin Evolution Diagrams showing loss in coherence as a transition, which has the slope corresponding to a transition onto QD\textsubscript{5D}, is crossed. The red dashed lines indicate these interdot transitions. The axes in the SEDs follow the same convention as in Figure~\ref{fig:figure1} of the main text. \textbf{e,} Spin Evolution Diagrams showing either a loss in coherence or change in g-factor susceptibility when a transition is crossed that, has the slope corresponding to a transition onto QD\textsubscript{2F}. The red dashed lines indicate these interdot transitions.}
    \label{fig:incoherent_dots}
\end{figure}

In the main manuscript, we demonstrated the ability to tune 30 out of 36 quantum dots in the $6\times6$ array through coherent probing using SEDs (Figure~\ref{fig:figure1}\textbf{c}). While the remaining quantum dot locations did not produce conclusive spin-coherent oscillations during the SED experiment, this doesn't exclude successful formation of the quantum dots at those sites. Here we look into features which are not based on spin-coherence, but can still signify their formation. We also speculate on the origin of the lack of coherence.\\

The quantum dots on the edge of the array are most readily found through direct charge sensing. During initial tune-up, all four SHTs around the array were tuned up during initial screening of the device, and four out of six remaining incoherent quantum dots were detected~(see Figure~\ref{fig:incoherent_dots}a-c). Since the charge-sensing data was collected only during the initial tune-up, no direct charge sensing data was gathered for the final two quantum dots QD\textsubscript{2F} and QD\textsubscript{5D}. As a result, we can only confidently claim successful formation of 34 out of 36 quantum dots. However, when looking at the transition lines in the SEDs~(see Figure~\ref{fig:incoherent_dots}d-e), interdot lines for both QD\textsubscript{2F} and QD\textsubscript{5D} are observed. We note that while coherence is observed on QD\textsubscript{2F}, inspecting the time-resolved oscillation does not give a well-defined frequency. We suspect that low tunnel coupling leads to latched shuttling hence labelling this site inaccessible for coherent shuttling. \\

Given our ability to discern the loading of all the remaining 6 inaccessible quantum dots we hypothesise that it is indeed possible to shuttle a charge into those quantum dots, and thus reaching all 36 out of 36 sites, albeit not always spin-coherently. If the incoherence results from the shuttling itself, these sites may still be used as static qubits. If the spin-information inherently decoheres at these sites, for example due to some material defect, they need to be avoided altogether by shuttling around them, which is possible given this sparse occupation. Moreover, a different potential landscape, attained through alternative tuning, may revive coherence at these sites as well. This however, is left for future work as retuning would affect currently functioning parts of the device.\\

It is difficult to decisively pinpoint the reason for the apparent lack of coherence when probing these six quantum dot sites, as the available statistics are limited. Generally, it may result from either incoherent shuttling or decoherence on the sites themselves. Incoherent shuttling due to low tunnel coupling has for example been observed in earlier devices as well~(Figure~\ref{fig:readout_features}b-c). Of these inaccessible quantum dots, five are located on the edge, and therefore a link could be made with various edge effects. Firstly, the change in the potential landscape could impact the shape of the quantum dots and the surrounding tunnel barriers. Moreover, the proximity to the single hole transistors and  ohmic reservoir might impact the spin-coherence at these sites. The relevance of this edge-effect pattern might be better understood in larger devices, in which there is a more prominent `bulk' region. Away from the edge, the inaccessible location QD\textsubscript{5D} could suffer from a lattice dislocation~\cite{stehouwerGermaniumWafersStrained2023} or some other material defect\AI{, as seen in previous germanium qubit devices in similar heterostructures~\cite{corley-wiciakNanoscaleMapping3D2023a}}. This is supported by a large DC voltage offset of $\text{P}_{5\text{D}}$ compared to the other plunger gates that is necessary to avoid the interdot transition onto $\text{QD}_{5\text{D}}$~(Figure~\ref{fig:incoherent_dots}d). Since the heterostructure is grown on a silicon wafer, \AI{and has an average threading dislocations density of $5\cdot10^{6}\,\SI{}{cm^{-2}}$~\cite{stehouwerGermaniumWafersStrained2023}, we expect on average to find one dislocation every $\SI{4}{\micro m^2}$, so} there is a non-negligible chance that one such dislocation is present in the $\SI{1.27}{\micro m^2}$ area spanned by the plunger gates. 
\section{Full Experimental Sequence}
\label{supp:ExperimentalSequence}

\begin{figure}
    \centering
    \includegraphics[width=\linewidth]{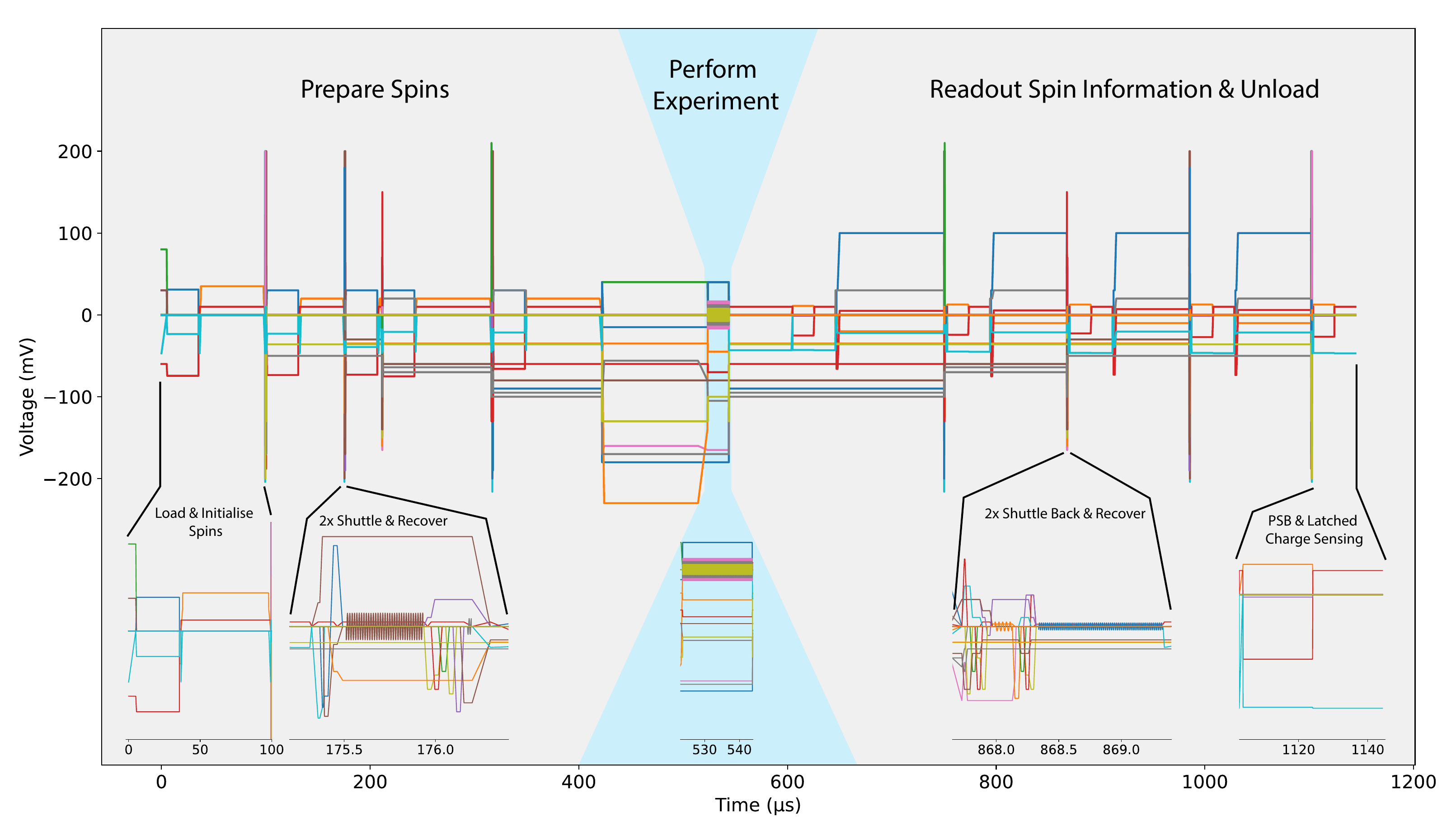}
    \caption{\textbf{Example Experimental Sequence Waveform.} A compiled waveform that is being played by the AWGs for some experiment is displayed. Both baseband pulses and microwave pulses are shown. The given baseband pulses correspond to virtual gates. Considering the variety of gates involved, the traces are not labelled. The sequence consists roughly of 3 parts. First, the 10 spins are prepared, in five sequential pairs, then the experiment of interest is performed. In this case, a simultaneous chevron experiment is run. Finally the spins are shuttled back, readout and unloaded. The insets show zoom-ins of particular parts of the sequence. Notice that the loading, initialisation and unloading take the most time, and the shuttling and experimental sequence itself are a relatively small fraction of the experimental time.}
    \label{fig:experimental_sequence}
\end{figure}
In the most complex experimental sequences of this manuscript, all ten qubit sites are initialised as basis states, after which the experiment is done and the qubit sites are read out. An experiment sequence in which all spins are loaded takes generally at least \SI{1}{ms}, with an example sequence given in Figure~\ref{fig:experimental_sequence}. In this section we will go through the exact steps that comprise the site-initialisation and spin-readout for the consecutive spin pairs. \\

To initialise any qubit, we need to load the charges first. We begin by making sure that the array, in particular the quantum dots around the reservoir are fully empty by purging site QD\textsubscript{1A-C} with a positive voltage pulse. Next we proceed to load exactly two charges into the array, first populating site QD\textsubscript{1B}. Due to a lack of fast pulsing lines, loading barrier B\textsubscript{l} has a fixed tunnel coupling and is set to neither strongly nor weakly couple the SHT with the array. Hence we need to wait several tens of microseconds before the two charges are loaded with high probability. We want to note, that this charge-step is the one most susceptible to voltage drift in the entire experimental sequence. Unlike the other transitions, which typically operate in an isolated regime, and therefore have a broad voltage window, the quantum dot here is connected to the reservoir, leading to additional spurious charge transitions. These spurious transitions need to be avoided by carefully aligning the energy levels of QD\textsubscript{1B} with the levels of the SHT. Typically, the AWG pulses involved require daily to weekly re-calibration. Interestingly, it seems that most voltage drift centres around the SHT, with for example the PSB window which will be discussed later being relatively robust against these voltage drifts.\\

Once the two desired charges are loaded, we change the vP\textsubscript{1B}-vP\textsubscript{1C} detuning to load the two charges to QD\textsubscript{1C}, and next the vP\textsubscript{1C}-vP\textsubscript{1D} detuning to move one of them to QD\textsubscript{1D}. We note that ramp times and wait-times are adjusted to assist with deterministic charge transfer. At this point we achieve the desired (1,1) charge configuration under QD\textsubscript{1C} and QD\textsubscript{1D} for us to start with spin-physics. \\

We initialise a well-defined spin state by increasing the P\textsubscript{1C}-P\textsubscript{1D} detuning and go to the (0,2) charge occupation. Here we wait for the spin state to decay from any orbital excitation to the single S(0,2) ground-state. By slowly ramping back across the QD\textsubscript{1B}-QD\textsubscript{1C} interdot transition, the spin-orbit anticrossing is passed adiabatically~\cite{kellyIdentifyingMitigatingErrors2025} and a $\ket{\downarrow\downarrow}$ ground state is initialised. Unlike during the charge-loading we are isolated from the reservoir and thus the interdot transition is extended. As a result, we are not sensitive to voltage drift, and do not need to recalibrate these voltages for several months.\\

\begin{figure}
    \centering
    \includegraphics[width=1\linewidth]{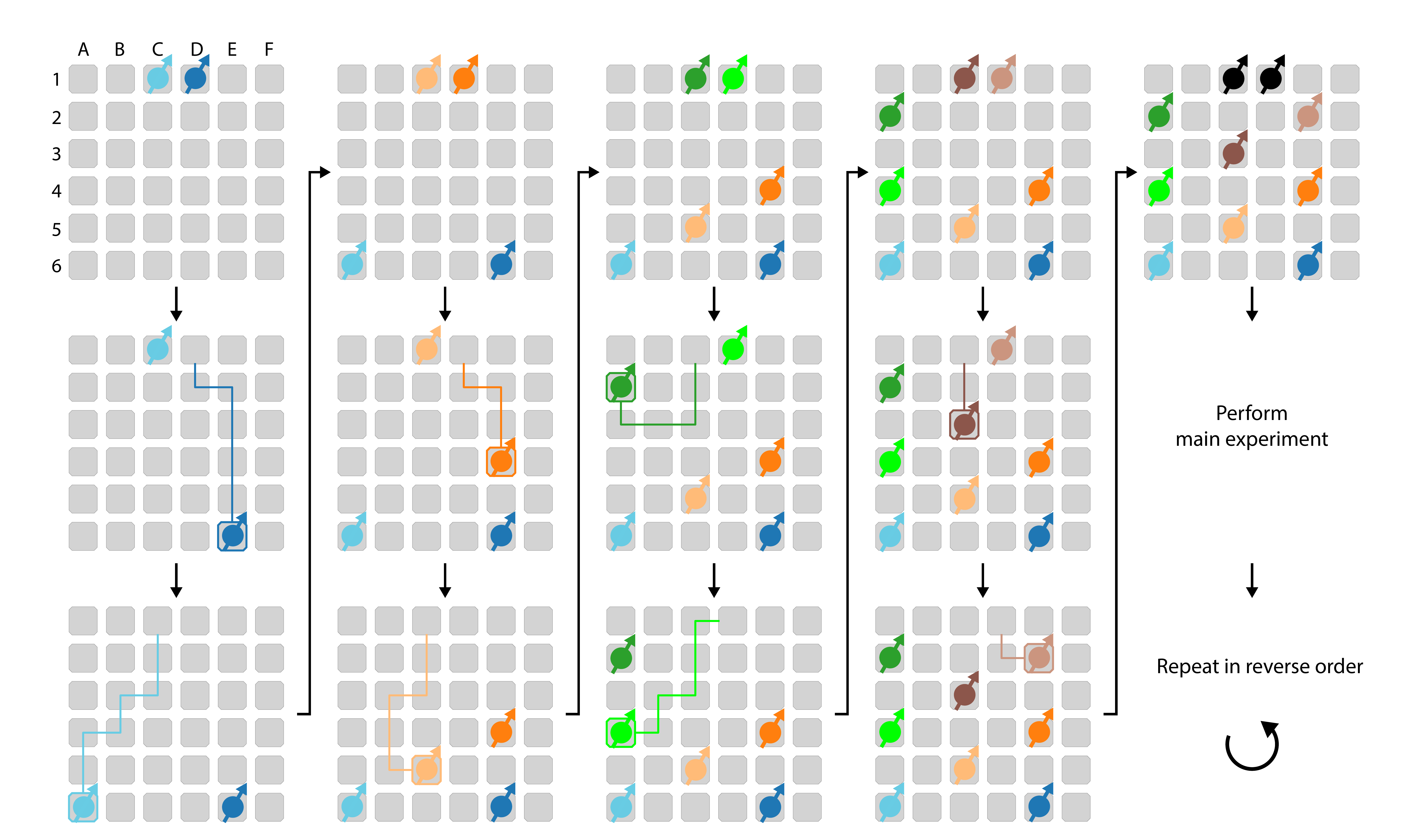}
    \caption{\textbf{Shuttling Steps.} An overview of all the paths the qubits are shuttled in and their order. Each time a pair of qubit appears in the array indicates when these are loaded. Spins that are loaded together are also used to adiabatically initialise the two-qubit ground state, and PSB is performed within that same pair. After the qubits are shuttled into position the experiment can be performed, and the qubits are shuttled back in reverse order for readout. Note that the recovery gates and reinitialisation operations are not indicated.}
    \label{fig:supp_shuttling_path_overview}
\end{figure}

Once spin states are redefined, we bucket-brigade shuttle them across the array to their desired position. The shuttling paths were visualised in Figure~\ref{fig:figure3} of the main text,  but are now indicated step by step in Figure~\ref{fig:supp_shuttling_path_overview}. As with the interdot transition used for initialisation, the voltages required for successful shuttling are robust to voltage drifts and have not been changed since initial tune-up. In addition to the robustness to drift, the extended interdots of the isolated operating regime also mean that gate-virtualisation is not as relevant. In particular, we do not take the effort of explicitly virtualising the plunger gate virtualisation within the array, and instead apply the gate-virtualisation also used at the edge of the array. We expect that in the future, these sparse architectures would not require any virtualisation at all, simplifying the required hardware, calibration steps, and software layers significantly.\\

Immediately after shuttling a spin to its desired location, a recovery gate~\cite{ademiDistributingEntanglementDistant2026} is performed to compensate for any shuttling induced rotation~\cite{vanriggelendoelman_2024_NatureCom_CoherentSpinQubit,Wang_2024_Science_OperatingSemiconductorQuantum,unseldBasebandControlSingleelectron2024,ademiDistributingEntanglementDistant2026}. Notably, this recovery gate is susceptible to the drift in Larmor frequency both along the shuttling path, as well as at the recovery location. As the shuttling itself is relatively fast compared to typical frequency drifts, it seems that the shift in the EDSR resonance has the most significant impact in our experiment. Therefore the driving frequency is typically calibrated daily, while the phase and duration of the recovery gate itself is recalibrated on the typical timescale of weeks following the procedure described in Supplementary Materials~\ref{supp:recovery_gates}. \\

After the recovery gate, the next spin of that pair is shuttled, or a new pair is loaded, until all 5 pairs are loaded and the experiment is executed. Once the experimental sequence is completed, the spin population is measured. This is done in reverse order to the one that the spins were loaded in, in other words: `Last In First Out'. So first the final, fifth, pair is measured through latching-enhanced PSB~\cite{harvey-collardHighFidelitySingleShotReadout2018,onoCurrentRectificationPauli2002}. The chosen PSB window is at the (1,1)-(0,2) interdot transition of the (QD\textsubscript{1C},QD\textsubscript{1D}) quantum dot pair. After the spin to charge conversion, depending on the spin state a charge will still be present in QD\textsubscript{1C} or not. This possible charge is moved to QD\textsubscript{1B} through bucket-brigade shuttling to a final, spin-dependent, charge occupation of (0,0,2) or (1,0,1) in the (QD\textsubscript{1B},QD\textsubscript{1C},QD\textsubscript{1D}) triple quantum dot system. This charge is read out using the top-left SHT, through reflectometry-based readout~\cite{Schoelkopf_RFSET_Science_1998}. Since the charge-occupation of QD\textsubscript{1B} may be in a meta-stable state, it is crucial that the tunnel-coupling between it and the reservoir is not too large. As this tunnel-coupling can not be easily dynamically tuned due to the lack of pulse-lines connected to B\textsubscript{l}, we choose an appropriate DC voltage, at the expense of slow loading and unloading of this quantum dot at other stages of the experiment. \\

After the spin-dependent charge is read out, a reference scan in the stable (0,1,1) is also taken, to compensate for any slow drift of the charge-sensor. Then, both spins are removed by following the loading steps in reverse with an additional voltage pulse on the array plunger gates to deplete them. These shuttling, readout and unloading steps are then sequentially repeated for all pairs to measure their spin state. 


\section{Recovery Gates and Shuttling}
\label{supp:recovery_gates}
In order to achieve high-fidelity initialisation and readout of the spin qubit array in the presence of diabatic shuttling-transitions, a recovery gate needs to be executed after each shuttling operation~\cite{ademiDistributingEntanglementDistant2026}. In the case of the current experiment, this requires two recovery gates per shuttled qubit, one after the outbound shuttling away from our initialisation sites (QD\textsubscript{1C} and QD\textsubscript{1D}) into the array, and one after the return trip before readout. Both recovery gates correct for a deterministic but random path-dependent unitary operation $U_{\text{shuttle}}$. It is convenient to rewrite this as an Euler decomposition $U_{\text{shuttle}}=R_\text{z}(\gamma)R_\text{x}(\beta)R_\text{z}(\alpha)$. Since the main purpose of the recovery gates in this manuscript is to initialise basis states and shuttle spin back for readout, the final phase added to the spin is not relevant and only $R_\text{z}(\gamma)R_\text{x}(\beta)$ has to be corrected for. In future experiments, the final phase $\alpha$ can be corrected through a virtual Z-rotation. Furthermore, since for qubit computation the distinction between $\ket{\uparrow}$ and $\ket{\downarrow}$ is a matter of relabelling (assuming two-qubit operations consist just of ZZ interactions), the recovery gate needs to compensate for the shuttling-induced rotation up to a spin-flip, i.e. $U_{\text{rec}}R_\text{z}(\gamma)R_\text{x}(\beta)=I\mod X$. As a result, a spin which was initially in the ground state may be flipped to the excited state after a perfect recovery gate, and vice versa. This flip is not detectable without the aid of another spin, or through $T_1$ relaxation events. In practice, recovery gate parameters are chosen to minimise the duration of the recovery gate. Still, to ensure proper readout, the outbound and return recovery gates must result in an overall identity, i.e. $U_{\text{rec,return}}R_\text{z}(\gamma_{\text{return}})R_\text{x}(\beta_{\text{return}})U_{\text{rec,out}}R_\text{z}(\gamma_{\text{out}})R_\text{x}(\beta_{\text{out}})=I$.\\

The recovery gates are implemented through electric dipole spin resonance (EDSR) driving, which means the unitary is a rotation around some axis given by $(\theta,\phi)$ in polar coordinates, for an amount $\Omega t_{\text{drive}}$, where $\Omega$ is the Rabi frequency and $t_{\text{drive}}$ is the drive time. $\theta$ is determined by detuning $\Delta = \omega_L-\omega_{MW}$ between the qubit Larmor frequency $\omega_L$ and the applied microwave frequency $\omega_{MW}$ as $\theta = \arctan(\Omega_0/\Delta)$, with $\Omega_0$ the on-resonance Rabi frequency. In our calibration steps, we generally fix the drive amplitude and the voltage landscape on top of which the drive is applied and therefore $\Omega_0$ remains fixed. We can change $\omega_{MW}, \phi$ and $t_{\text{drive}}$ in order to generate any arbitrary unitary rotation. Since EDSR driving typically has the highest fidelity when driving on resonance, resonant driving is assumed and therefore $\theta=\pi/2$. We note, however, that some small detuning can often be compensated with the appropriate drive-time and phase without significant complications. For simplicity, we'll assume that resonant EDSR driving is achieved, performing the rotation $R_{\phi}(\Omega t_{drive})$ around the equator of the Bloch sphere at a phase $\phi$ and angle $\Omega t_{drive}$. Therefore the goal of a successful recovery gate calibration is to find a $\phi$ and $t_{\text{drive}}$ such that $R_{\phi}(\Omega t_{drive})R_\text{z}(\gamma)R_\text{x}(\beta)=I\mod X$, which is solved by $\phi = -\gamma + n\pi$ and $t_{\text{drive}}=((1-(-1)^n)\pi/2-(-1)^n\beta\mod\pi)/\Omega$ for $n\in\mathbb{Z}$.\\
\begin{figure}
    \centering
    \includegraphics[width=\linewidth]{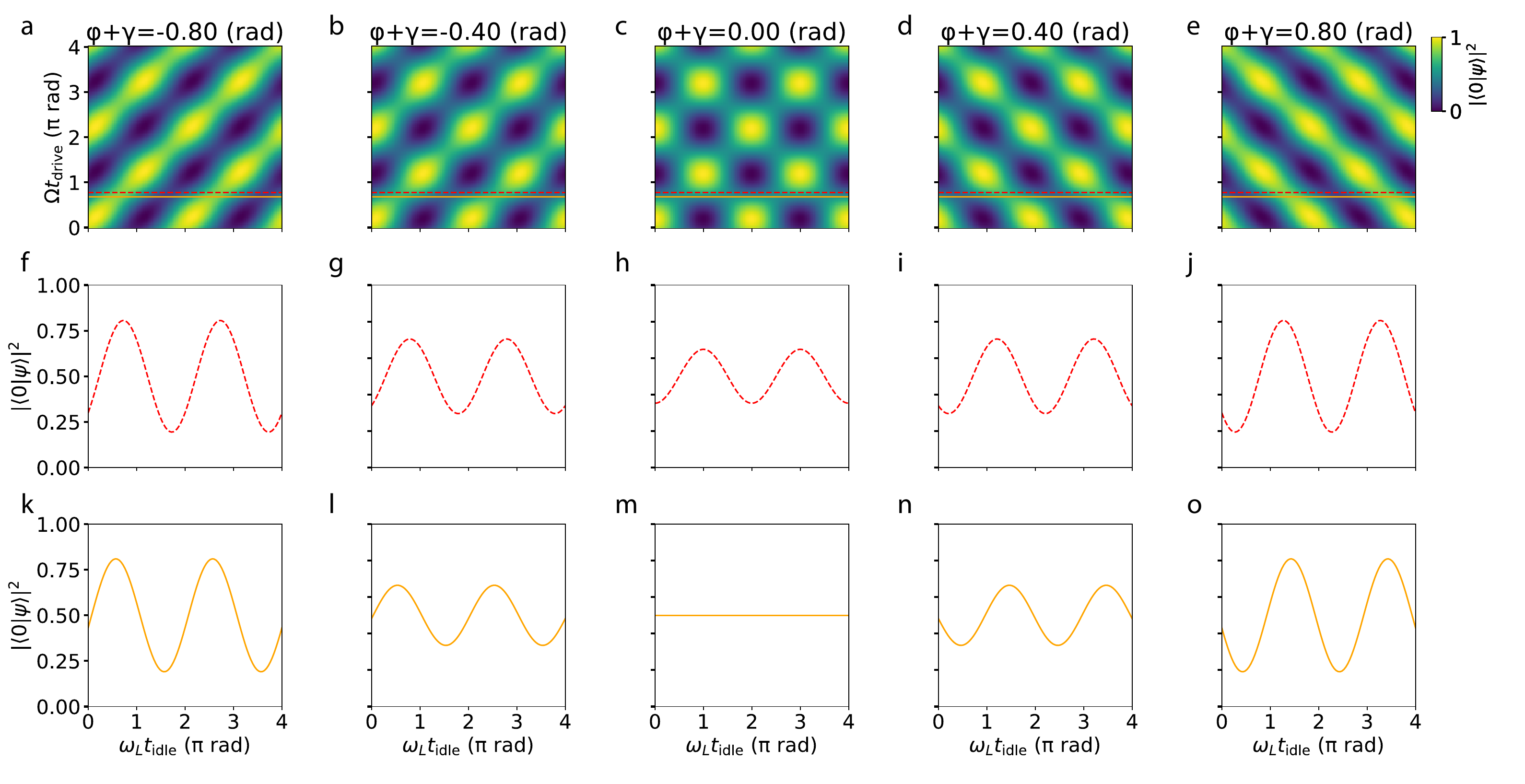}
    \caption{\textbf{Simulated Recovery Gate Calibration.} \textbf{a-e,} The simulated state population of a superposition state $\ket{+}$ after free evolution $R_\text{z}(\omega_L t_\text{idle})$, shuttling-induced operation $R_\text{x}(\beta)R_\text{z}(\gamma)$ and a resonant EDSR recovery gate $R_{\phi}(\Omega t_{\text{drive}})$ are applied, for different values of $\phi$ relative to $\gamma$. The shuttling-induced transversal rotation is fixed at $\beta=1$ rad. Several linecuts that correspond to different drive angles $\Omega t_\text{drive}$ of the recovery gate are given, indicating overdriving (dashed) and well-calibrated driving (solid). \textbf{f-j,} Overdriving produces oscillations as a function of the free evolution for any phase $\phi$ of the EDSR pulse. \textbf{k-o} Whenever the drive angle of the EDSR is $\Omega t_\text{drive}=\pi-\beta$, there is a phase $\phi=-\gamma$ (see \textbf{m}) at which no free evolution is observed. At these parameters the recovery gate calibration is successful.}
    \label{fig:recovery_gate_calibration}
\end{figure}

Our recovery gate calibration procedure consists of two main steps. We start by calibrating the return recovery gate first, after which the outbound recovery gate is calibrated. This removes Larmor oscillations in the distant quantum dot when calibrating the outbound recovery gate. For the return recovery gate, we first find the qubits' resonance frequency in the return quantum dot. Since the spin may start that scan in an arbitrary superposition, any frequency sweep with a given drive time may be skewed and asymmetrical. This makes typical methods more involved. We used the chevron data directly to estimate the desired frequency initially by eye. Once both the recovery gates are roughly calibrated, the frequency could be estimated again through standard methods, if so desired. \\

As noted, exact calibration of the frequency is less crucial than the subsequent phase and drive-time calibration. To calibrate these, we need to probe $\gamma$ and $\beta$, for which we induce free induction oscillations in the distant destination dot as a function of idle time $t_{\text{idle}}$. Then we sweep $\phi$ and $t_{\text{drive}}$ of the recovery gate and read out the spin population. At the drive parameters for which the spin dynamics are imperceptible, the recovery gate has been properly calibrated. A simulation of different drive parameters is given in Figure~\ref{fig:recovery_gate_calibration}. 
The benefit of this method is that the amplitude of the observed oscillations is sensitive to first order to any change in $t_{\text{drive}}$ and $\phi_{\text{drive}}$, unlike a procedure in which the basis-state population is subject to maximisation, where it is sensitive to only second order. We also stress that it is not trivial to amplify the error of such recovery gate-calibration through repeated shuttling-operation as is typically done~\cite{ademiDistributingEntanglementDistant2026}, because the total sequence would consist of two recovery gates. Thus, in a repeated procedure, any miscalibrations in one gate may be compensated by the other, even if neither of them is the correct recovery gate for their respective shuttling operation. This could be solved by letting the state dephase between each recovery gate, at the cost of potentially reduced visibility. This `decoherence stabilised' error amplification has not been studied or used in the current manuscript.\\

Once the return recovery calibration is complete, the outbound recovery gate can be calibrated in a comparable manner. Again, we find the frequency using the chevron, and let the spin idle after the execution of the outbound recovery gate, while sweeping the phase and duration of this gate. For readout in the origin site we measure the phase of the spin instead of the polarisation by inserting an X\textsubscript{90} before PSB. \\

\begin{figure}
    \centering
    \includegraphics[width=0.8\linewidth]{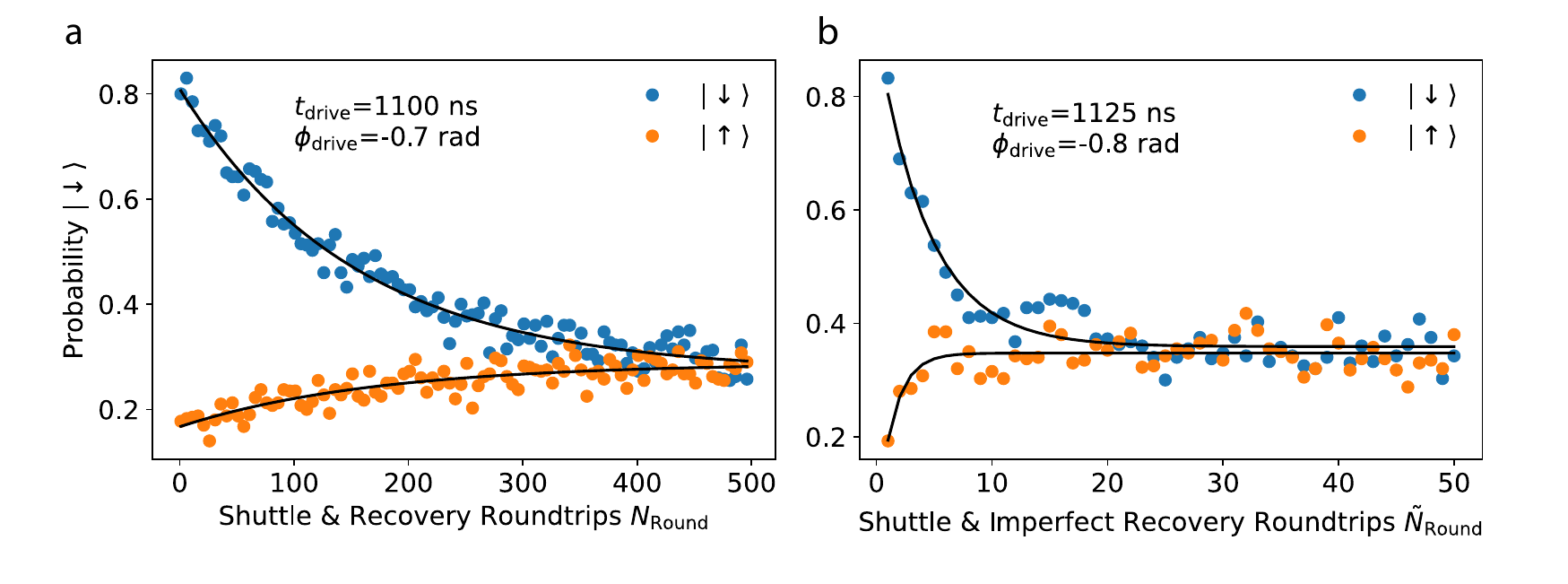}
    \caption{\textbf{Impact of miscalibrated recovery gate.} \textbf{a,} The recovery gate for the shuttling roundtrip between  QD\textsubscript{1C}-QD\textsubscript{3C}, with a fidelity of $F_\text{Round}=99.3\%$, using a recovery gate in QD\textsubscript{303} with drive time $t_\text{drive}=\SI{1100}{ns}$ and drive phase of $\phi=-0.7$\,rad. The fidelity is measured by looking at the basis-state decay as a function of the number of back-and-forth shuttles $N_{\text{Round}}$. \textbf{b,} Slightly changing the drive time and phase to $t_\text{drive}=\SI{1125}{ns}$ and $\phi=-0.8$\,rad significantly diminishes the roundtrip fidelity to $\tilde{F}_{\text{Round}}=75.8\%$.}
    \label{fig:miscalibrated_recovery}
\end{figure}

We note that imperfections in the recovery gate are likely a major source of infidelities in some shuttling paths, as exemplified by the fact that shuttling between QD\textsubscript{1D} and QD\textsubscript{4E} has a worse fidelity than between QD\textsubscript{1D} and QD\textsubscript{6E} even though the latter traverses the same path. In fact, we notice that small changes in the recovery gate may already result in a large drop in fidelity, as seen in Figure~\ref{fig:miscalibrated_recovery}. As a result, given the long experimental time required for the calibration, we simply may not have reached the optimal parameter regime. Apart from miscalibration, other coherent errors may include time-dependence due to heating~\cite{undsethHotterEasierUnexpected2023} or the disturbance of the other spin necessary for PSB readout during the shuttling of a given spin.

\section{Randomised Benchmarking Detailed Results}
\label{supp:1Q}
\begin{figure}
    \centering
    \includegraphics[width=1\linewidth]{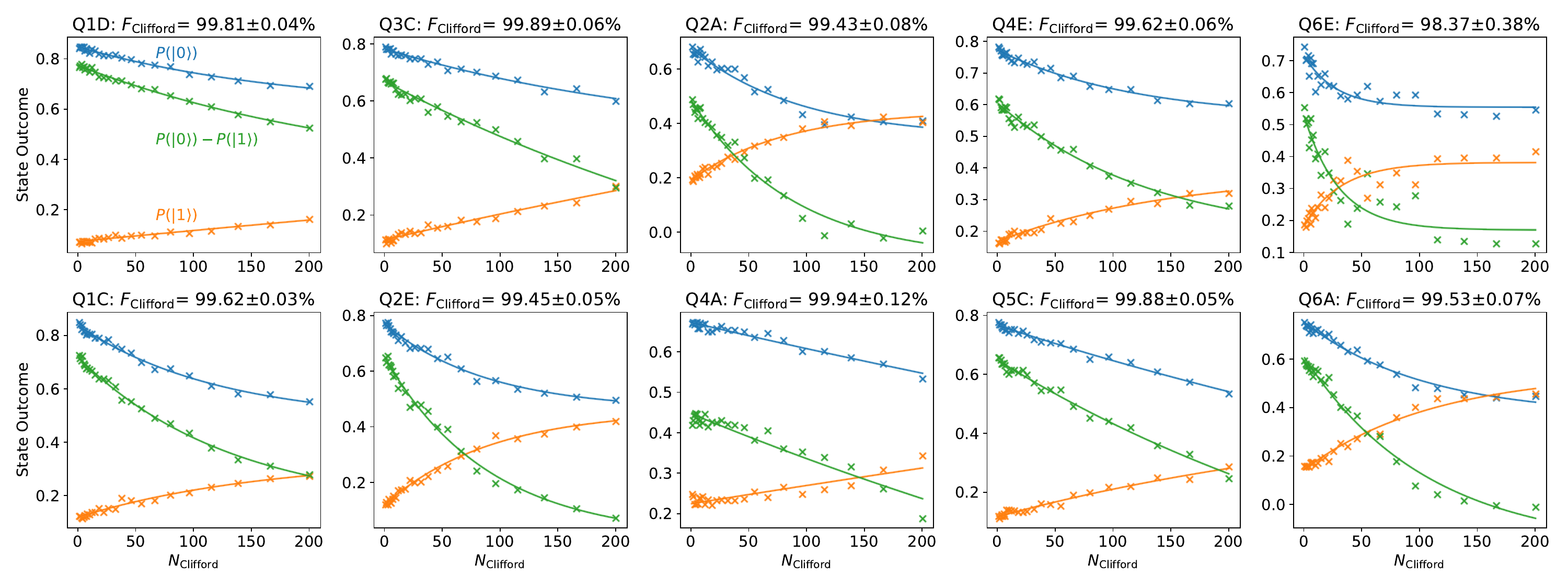}
    \caption{\textbf{Individual Randomised Benchmarking.} The decay curves for the individually performed RB are shown for each of the ten qubits. The curves give the results for the $\ket{0}$ and $\ket{1}$ probabilities, as well as their difference. The fidelity is extracted from the decay of this difference.}
    \label{fig:supp_1Q_RB_Ind}
\end{figure}

\begin{figure}
    \centering
    \includegraphics[width=1\linewidth]{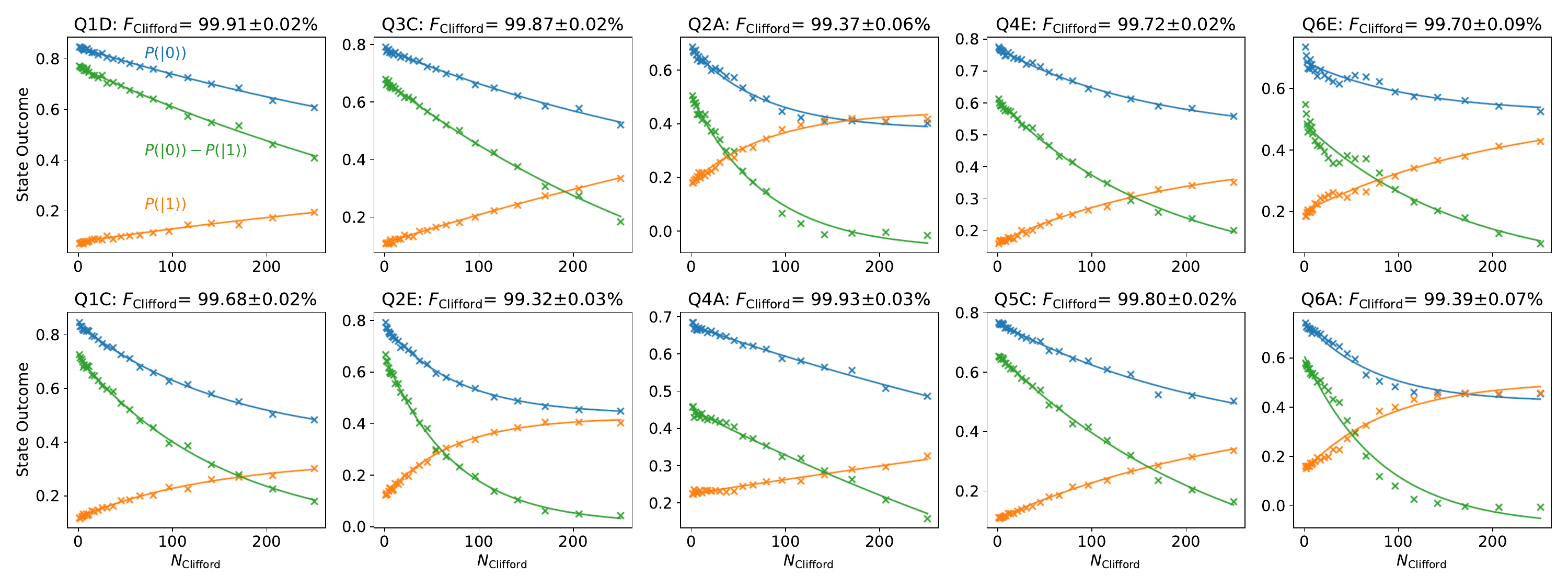}
    \caption{\textbf{Simultaneous Randomised Benchmarking.} As with the individual RB results, the decay curves for the simultaneously performed RB are shown for each of the ten qubits. The curves give the results for the $\ket{0}$ and $\ket{1}$ probabilities, as well as their difference. The fidelity is extracted from the decay of this difference. Within each row of qubits the measurements are taken from the same dataset, with the qubits in the other row acting as readout ancillas. }
    \label{fig:supp_1Q_RB_Par}
\end{figure}

\begin{figure}
    \centering
    \includegraphics[width=\linewidth]{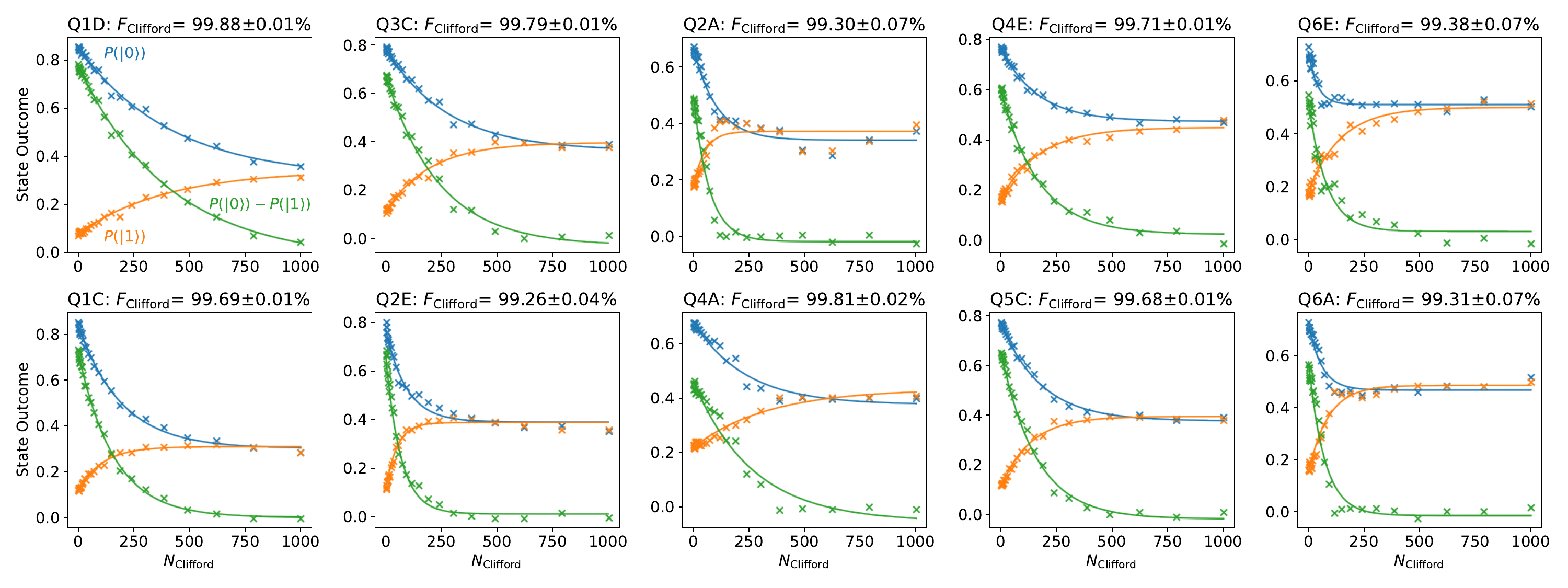}
    \caption{\textbf{Extended Simultaneous Randomised Benchmarking.} Same experiment as in Figure~\ref{fig:supp_1Q_RB_Par}, but over a longer range of  Clifford gates to observe saturation. The results are confirmed to be comparable.}
    \label{fig:supp_1Q_RB_Par_Long}
\end{figure}

In this section, we present the decay-curves from which the results in the main Figure~\ref{fig:figure4}e were extracted. The results for the individual and parallel Randomised Benchmarking (RB) are seen in Figures~\ref{fig:supp_1Q_RB_Ind} and~\ref{fig:supp_1Q_RB_Par}, respectively. In addition, the simultaneous RB has been executed up to $N_{\text{Clifford}}=1000$ to confirm the negligible impact of slightly different time durations of the sequence (Figure~\ref{fig:supp_1Q_RB_Par_Long}). For the main, the shorter sequence has been used for a fairer comparison with the individual 1Q RB, which has been performed up to $N_{\text{Clifford}}=200$ to avoid the need for intermediate re-calibration given the long experimental time required. 

\section{Two Qubit interaction}
\label{supp:2Q}
In this section, we provide further details about the two qubit interactions studied in the manuscript. First, we present the time-resolved data for $q_{\text{4A}}-q_{\text{5C}}$ interaction, after which the effect of adding decoupling pulses will be examined. Finally, details on the exchange interaction for other pairs will be discussed.
\subsection{Time Resolved Interaction Q\textsubscript{4A}-Q\textsubscript{5C}}
\begin{figure}
    \centering
    \includegraphics[width=\linewidth]{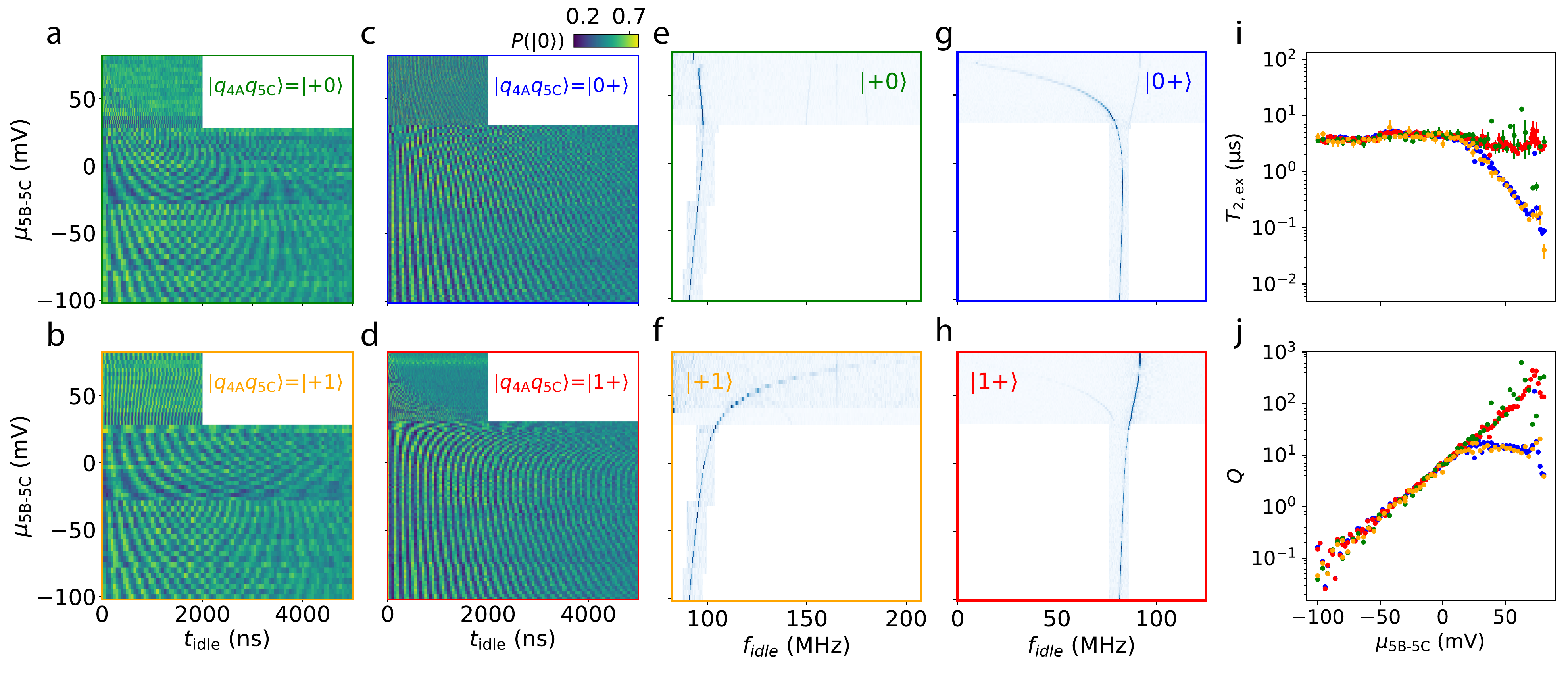}
    \caption{\textbf{Time-resolved exchange oscillations.} \textbf{a-d,} The time-resolved Larmor oscillations during the interaction between $q_{\text{4A}}$ and $q_{\text{5C}}$. The state that was initialised in this Ramsey-like experiment is stated in the inset. \textbf{e-h,} The Fourier Transform of the time-resolved oscillations for each initial state. This is the same data which is combined in main Figure~\ref{fig:figure5}c. \textbf{i,} The characteristic decay time of the observed exchange oscillation for each state, coloured according to the initial state like in the other plots. \textbf{j,} Quality factor $Q=J\cdot T_{\text{2,ex}}$ of the exchange oscillations. Here $J$ is extracted from the spectrum, and can be seen in main Figure~\ref{fig:figure5}d.}
    \label{fig:time_resolved}
\end{figure}
Since the interaction strength $J$ varies over three orders of magnitude (see main Figure~\ref{fig:figure5}d), the time-resolved data was split into several regions. This allows us to pick a minimal time resolution and range within the relevant bounds for that region to speed up the data collection. The raw time-resolved data of the exchange oscillations as a function of $\mu_{\text{5B-5C}}$, at $\varepsilon_{\text{5B-5C}}=0$ is presented in Figure~\ref{fig:time_resolved}. The discontinuities in this figure stem from different artificial detuning values used in the different regions, to allow for an easier fit. The observed data is fitted to an exponentially decaying oscillation, whose frequency and decay time are presented in Figure~\ref{fig:figure5}c,d of the main text.  
\subsection{Character Randomised Benchmarking Q\textsubscript{4A}-Q\textsubscript{5C}}
\begin{figure}
    \centering
    \includegraphics[width=0.7\linewidth]{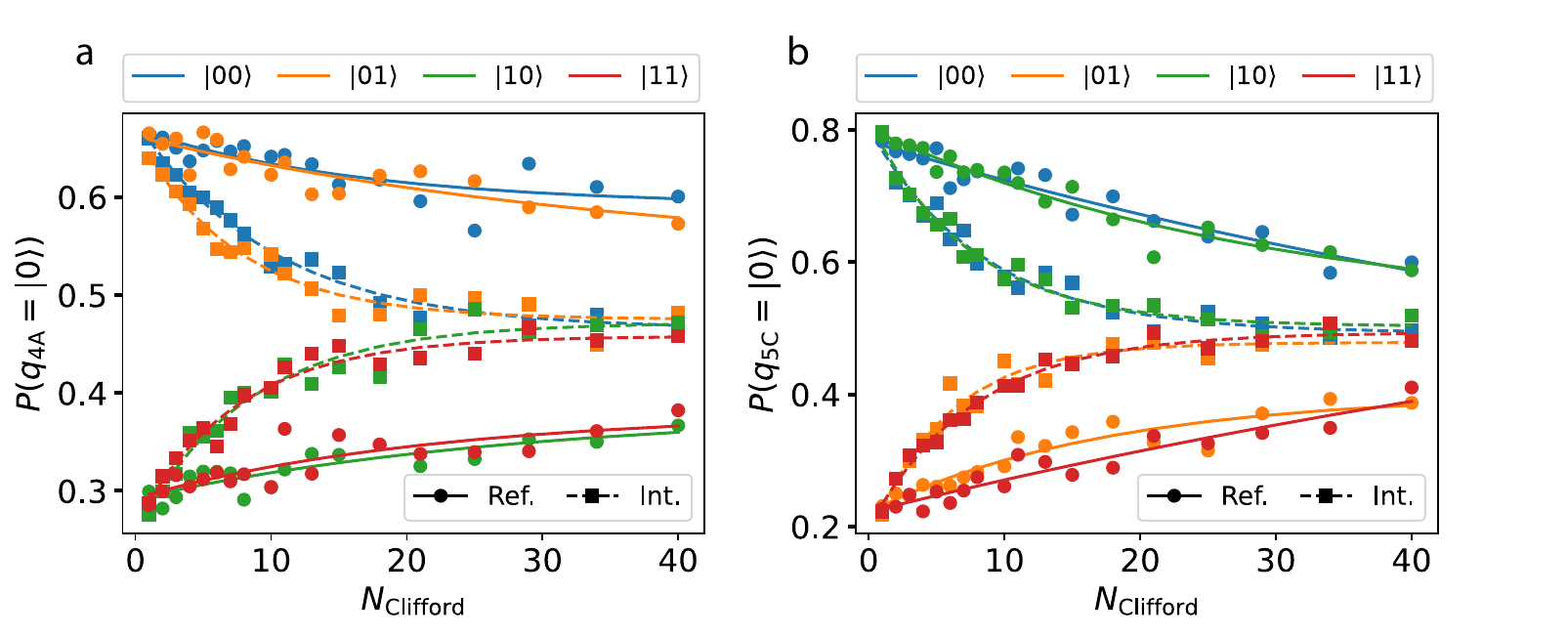}
    \caption{\textbf{Character Randomised Benchmarking Data.} \textbf{a} and \textbf{b} show the probability of measuring respectively $q_{\text{4A}}$ and $q_{\text{5C}}$ to be in the $\ket{0}$ state, for different fiducial states $\ket{\psi_{\text{init}}}=\ket{q_{\text{4A}}q_{\text{5C}}}$ of the character randomised benchmarking sequence. Both the reference (Ref.) and CZ-interleaved (Int.) sequences have been taken for $N_{\text{random}}=10$ randomisations.}
    \label{fig:detailed_CRB}
\end{figure}
In the main text, we presented an averaged Character Randomised Benchmarking dataset to showcase the 2Q gate fidelity. This is only possible with minimal assumptions if the data is unbiased. Indeed, we see in Figure~\ref{fig:detailed_CRB} that the raw data is symmetric around its decohered value for the different CRB fiducials used. This allows us to average the data after rectification of the relevant fiducial states. 
\subsection{Decoupled CZ}
\begin{figure}
    \centering
    \includegraphics[width=0.8\linewidth]{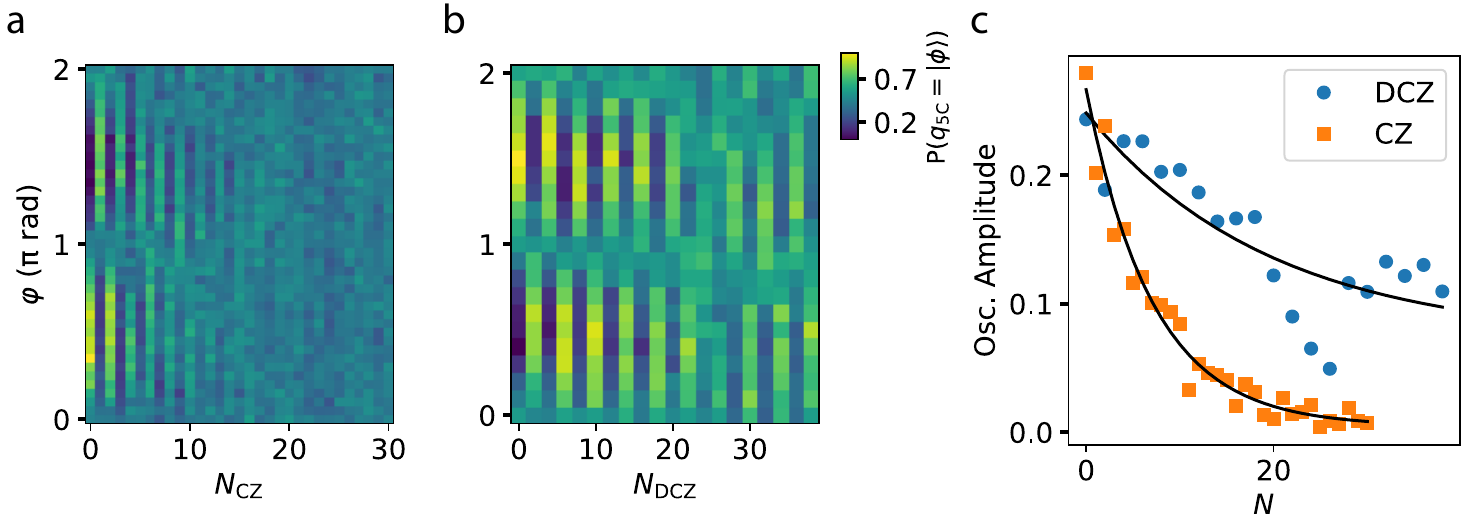}
    \caption{\textbf{Repeated Exchange gates.} The state $q_{\text{4A}}q_{\text{5C}}=\ket{1}\ket{+i}$ dephases after multiple applications of respectively the CZ \textbf{(a)} and DCZ \textbf{(b)} gate. To show the coherence and phase evolution, the probability of $q_{\text{5C}}$ is displayed as projected on $\ket{\phi}=\ket{0}+e^{-i\phi}\ket{1}$. Note that due to its structure, the DCZ gate is four-periodic instead of two-periodic like the regular CZ. \textbf{c,} Decay of the superposition state oscillation amplitude as extracted from \textbf{a} and \textbf{b}. The black line is an exponential fit of the data points.}
    \label{fig:repeated_CZ}
\end{figure}
Since the CZ gate includes shuttling the spins towards each other, which may need to be slow to ensure spin-adiabaticity, the quantum state may decohere during the shuttling process. To study the impact of the dephasing time $T_2^*$ on the quality of the CZ gate, we introduce a double-decoupling pulse $X_{\pi,q\text{4A}}\otimes X_{\pi,q\text{5C}}$ to cancel out quasi-static noise on each qubit, splitting up the CZ into a decoupled CZ (DCZ) gate~\cite{watson_2018_Nature_ProgrammableTwoqubitQuantum}. To simplify calibration and phase-tracking, the decoupling pulses are executed in the separated quantum dots (as in main Figure~\ref{fig:figure5}a). This does lead to increased shuttling, as the spins need to be shuttled back to their separate quantum dots to perform the echo. Therefore, the decoupling pulses are expected to only result in improved gate-quality if the coherence enhancement outweighs the time increase. Still, we see in Figure~\ref{fig:repeated_CZ} that indeed the superposition state ${q_{\text{5C}}}_{\text{init}}=\ket{+}$ retains coherence for longer when it is subject to the DCZ gate. Analysing the decay in Figure~\ref{fig:repeated_CZ}, a $1/e$ decay is observed after $N_{\text{DCZ}}^*=19$ DCZ repetitions compared to $N_{\text{CZ}}^*=7$ repetitions for the CZ. The enhancement observed from the decoupling strengthens the idea that improving the coherence time would result in higher-quality two-qubit gates, and is hence paramount in future work. This is particularly necessary for the qubit pairs where shuttling needs to be relatively slow to ensure adiabaticity.
\subsection{Other interacting qubits}
\begin{figure}
    \centering
    \includegraphics[width=0.9\linewidth]{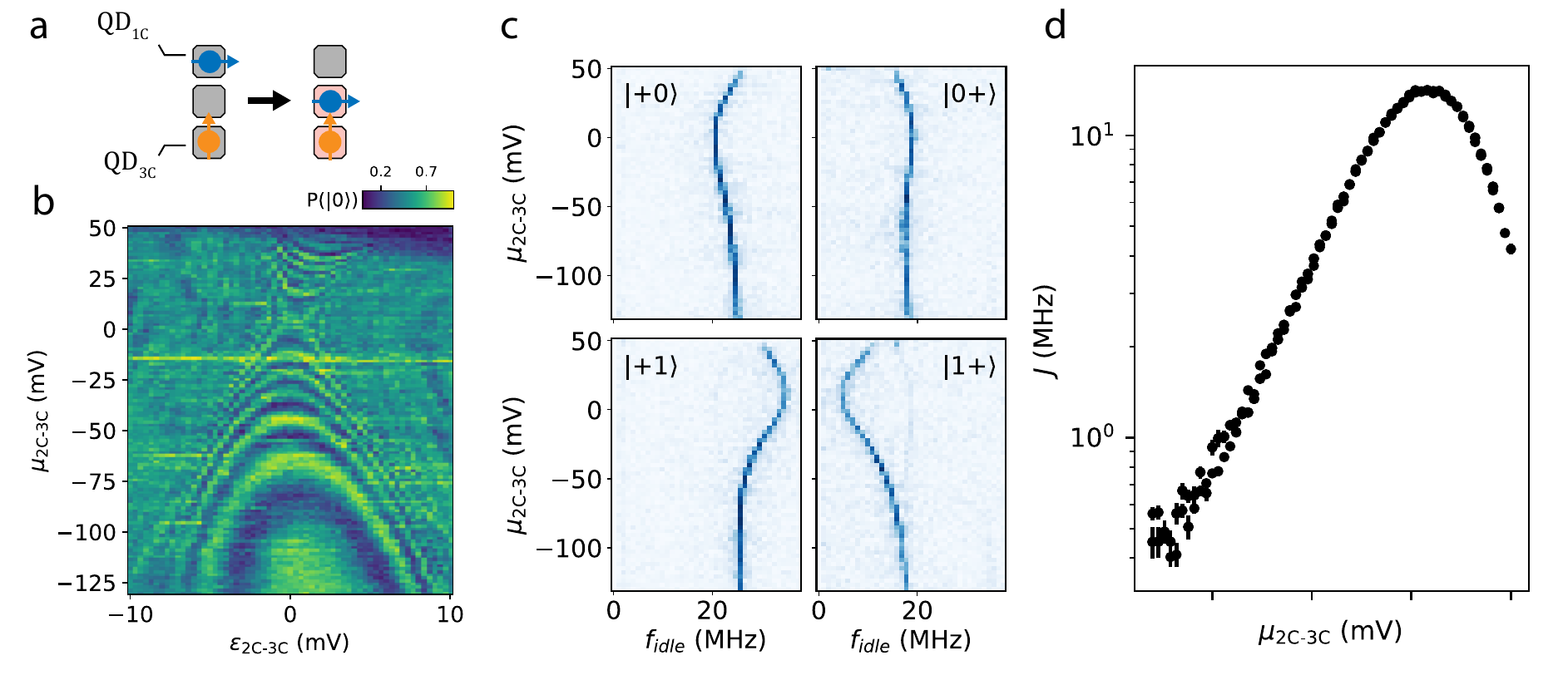}
    \caption{\textbf{2Q Interaction of }$\bm{q}_{\text{\textbf{1C}}}$\textbf{-}$\bm{q}_{\text{\textbf{3C}}}$\textbf{.} \textbf{a,} Schematic of the plunger gates where the spins reside under and where they are shuttled towards for the interaction. The 2Q interaction is tuned at the location indicated by the pink plunger gates. The horizontally oriented spin is used as a probe for the decoupled fingerprint measurement in \textbf{b}. \textbf{c,} The idling oscillation frequency for different initial states of $\ket{q_{1\text{C}}q_{3\text{C}}}$ from which the exchange interaction in \textbf{d} is computed. This frequency is extracted along the $\varepsilon_{\textbf{2C-3C}}=0$ detuning line in the fingerprint plot of \textbf{b}.}
    \label{fig:2Q_103_303}
\end{figure}
Next to pair $q_{\text{4A}}$-$q_{\text{5C}}$, we manage to connect three other qubit pairs through exchange interaction: $q_{\text{1C}}$-$q_{\text{3C}}$, $q_{\text{4A}}$-$q_{\text{6A}}$ and $q_{\text{3C}}$-$q_{\text{4A}}$. This way, a connected graph can be made between qubits from each of the five loaded qubit pairs~(see Figure~\ref{fig:supp_shuttling_path_overview} for the pair designations). More connections have not been established, largely due to coherence loss for certain links, for example the inaccessibility of QD\textsubscript{5D} or the low fidelity connection between QD\textsubscript{3C}-QD\textsubscript{4C} (see the SEDs in main Figure~\ref{fig:figure1}c). This coherence may be improved with additional tuning, but this remains to be studied. Moreover, increasing the sparsity may enable greater flexibility in chosen shuttling paths once the array is occupied with qubits. \\
\begin{figure}
    \centering
    \includegraphics[width=0.9\linewidth]{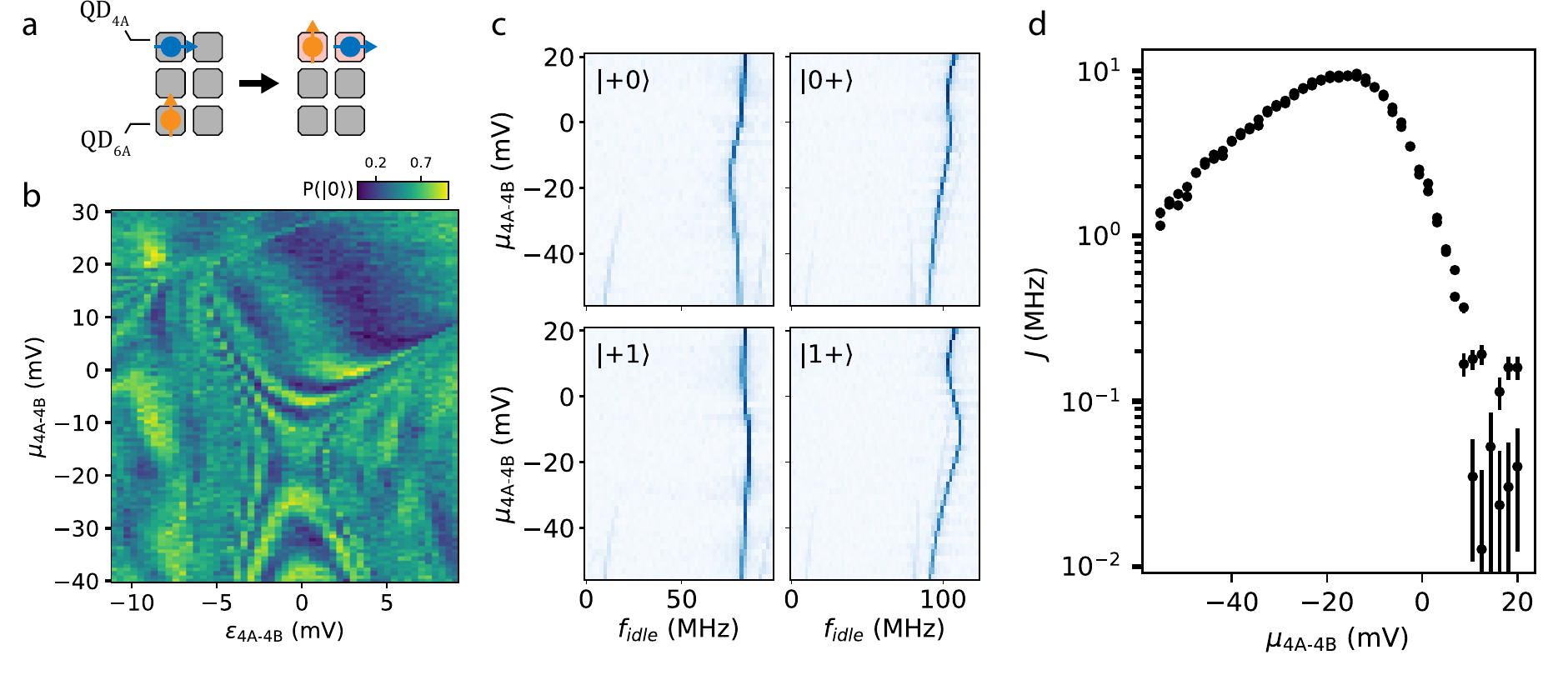}
    \caption{\textbf{2Q Interaction of }$\bm{q}_{\text{\textbf{4A}}}$\textbf{-}$\bm{q}_{\text{\textbf{6A}}}$\textbf{.} \textbf{a,} Schematic of the plunger gates where the spins reside under and where they are shuttled towards for the interaction. The 2Q interaction is tuned at the location indicated by the pink plunger gates. The horizontally oriented spin is used as a probe for the decoupled fingerprint measurement in \textbf{b}. \textbf{c,} The idling oscillation frequency for different initial states of $\ket{q_{4\text{A}}q_{6\text{A}}}$ from which the exchange interaction in \textbf{d} is computed. This frequency is extracted along the $\varepsilon_{\textbf{4A-4B}}=0$ detuning line in the fingerprint plot of \textbf{b}.}
    \label{fig:2Q_401_601}
\end{figure}

\begin{figure}
    \centering
    \includegraphics[width=0.9\linewidth]{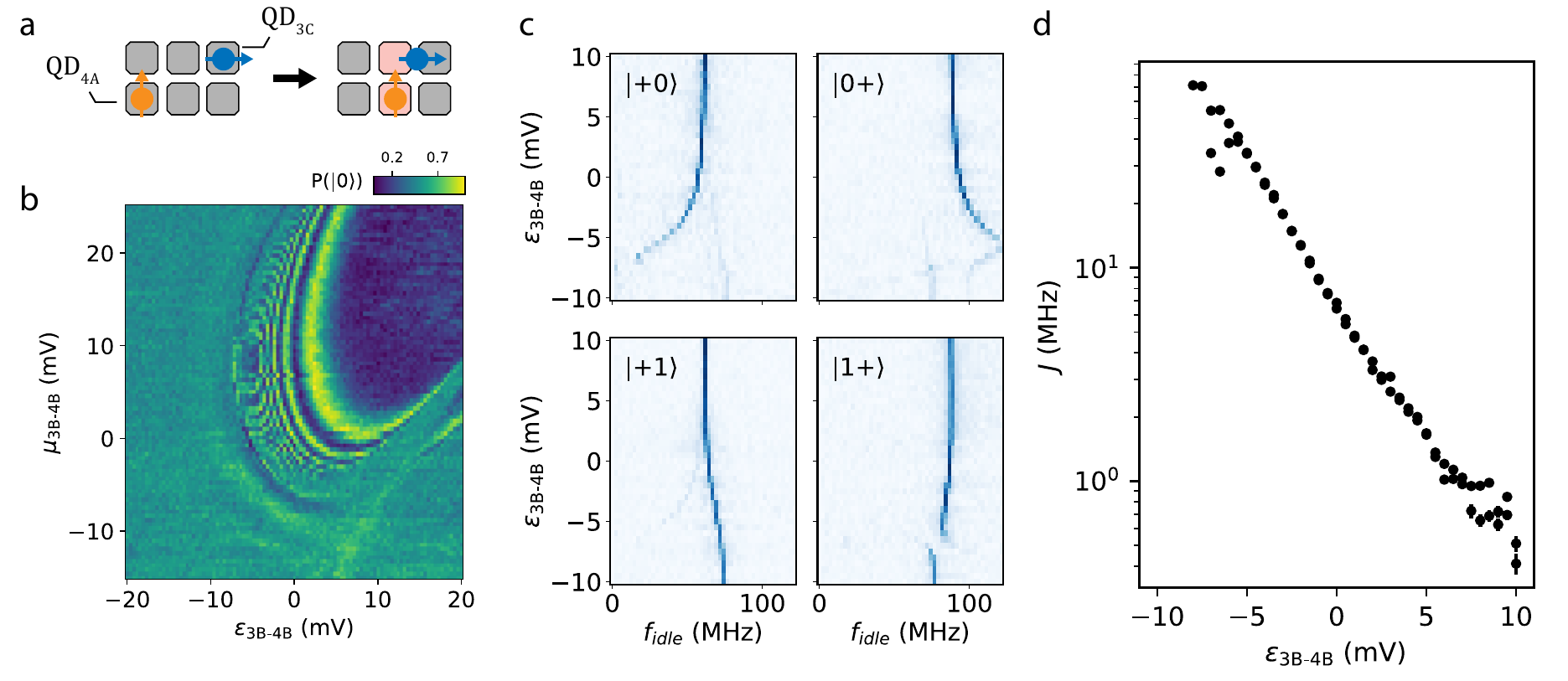}
    \caption{\textbf{2Q Interaction of }$\bm{q}_{\text{\textbf{3C}}}$\textbf{-}$\bm{q}_{\text{\textbf{4A}}}$\textbf{.} \textbf{a,} Schematic of the plunger gates where the spins reside under and where they are shuttled towards for the interaction. The 2Q interaction is tuned at the location indicated by the pink plunger gates. The horizontally oriented spin is used as a probe for the decoupled fingerprint measurement in \textbf{b}. \textbf{c,} The idling oscillation frequency for different initial states of $\ket{q_{3\text{C}}q_{4\text{A}}}$ from which the exchange interaction in \textbf{d} is computed. This frequency is extracted along the $\mu_{\textbf{3B-4B}}=5$ line in the fingerprint plot of \textbf{b}. The exchange energy is extracted up to a point where the fit has high enough quality, and extends to the low end of $\varepsilon_{\text{3B-4B}}$.}
    \label{fig:2Q_401_303}
\end{figure}

For the interactions that can be established, the shuttling paths, fingerprint plots, oscillation frequencies and exchange interactions are extracted. For these pairs we observe more complicated behaviour than for the qubit pair presented in the main text. Considering $q_{\text{1C}}$-$q_{\text{3C}}$, one sees that the typical fingerprint plot gives way to a more unusual hourglass-like structure~(Figure~\ref{fig:2Q_103_303}b). Looking at the qubit precession, we indeed see that a maximum is reached as a function of energy $\mu$ at the waist of the hourglass, after which the exchange interaction decreases again. We hypothesise that this behaviour is related to the electrostatic landscape outside the double quantum dot picture, and could be explained by involving a third quantum dot. In particular, we suspect that $q_{\text{1C}}$ is delocalised across more quantum dots, presumably QD\textsubscript{2C} and QD\textsubscript{1C}. Given the large parameter space that impacts the electrostatic potential of the quantum dots, and the interactions arising from the presence of many spins in the array, in-depth study of these phenomena is left for future work. For this spin-pair, we manage to tune the exchange interaction by over an order of magnitude by using symmetric plunger tuning, being limited from below by the AWG range and from above by this hourglass behaviour. A similar behaviour is observed for the $q_{\text{4A}}$-$q_{\text{6A}}$ pair (Figure~\ref{fig:2Q_401_601}), where the AWG range is even more limiting in the current DC tuning configuration. For $q_{\text{3C}}$-$q_{\text{4A}}$, the departure from the expected picture is more stark, and we expect that the spins do not move fully into neighbouring quantum dots~(Figure~\ref{fig:2Q_401_303}). Instead, we suspect that the plunger gate acts more like a conventional barrier gate, modulating the tunnel barrier between the two spins instead of modulating the depth of the potential well in which the spins reside. Investigating the parameter space more broadly was again challenging at this DC tuning, due to the limitations of the AWG voltage range on the one hand and the impact on the other spins of the array on the other.\\ 

Altogether, these dynamics show that it will be crucial in future experimental work to better comprehend the energy landscape in which multi-spin interactions take place, and how it can be controlled more reliably. We also encourage theoretical work simulating two-spin interactions as they are being shuttled between different quantum dots towards one another, since shuttling-mediated two-qubit interactions are likely to be a key element of future spin qubit architectures. In terms of CZ-gate fidelity, Character Randomised Benchmarking results reveal fidelities of $71.5\pm2.5\%$, $68.4\pm2.4\%$ and $66.8\pm6.0\%$ for the QD\textsubscript{1C}-QD\textsubscript{3C}, QD\textsubscript{4A}-QD\textsubscript{6A} and QD\textsubscript{4A}-QD\textsubscript{3C} pairs respectively. These fidelities are understood to be largely limited by slow shuttling or residual diabatic transitions, as well as, in the case of the QD\textsubscript{4A}-QD\textsubscript{3C} pair in particular, a noise-sensitive wavefunction configuration.\\

